\documentclass[11pt,tightenlines,eqsecnum,floats,aps,amsmath,amssymb,nofootinbib,prd,shownopacs]{revtex4}
\usepackage{amsmath,amssymb,amsfonts,amsthm}
\usepackage{graphicx}
\usepackage{enumerate} 
\usepackage{colordvi} 

\usepackage[mathscr]{eucal}
\usepackage[latin1]{inputenc}
\usepackage{amsmath}
\usepackage{amsfonts}
\usepackage{amssymb}
\usepackage{graphicx}

\def\lp{{\ell}_{\rm Pl}}

\def\q{\mathring{q}}
\def\e{\mathring{e}}
\def\w{\mathring{\omega}}
\newcommand{\norm}[1]{\ensuremath{\|#1\|}}

\newcommand{\rcr}{\rho_{\mathrm{crit}}}

\newcommand{\p}{\partial}

\newcommand{\f}{\frac}

\def\rmax{\rho_{\mathrm{max}}}

\def\rcr{\rmax}

\usepackage{enumerate}

\def\f{\frac}

\def\d{\textrm{d}}

\def\ul{\underline}

\usepackage{enumerate}

\newcommand{\be}{\nopagebreak[3]\begin{equation}}
\newcommand{\ee}{\end{equation}}
\newcommand{\ba}{\nopagebreak[3]\begin{eqnarray}}
\newcommand{\ea}{\end{eqnarray}}

\newcommand{\bmult}{\nopagebreak[3]\begin{multline}}
\newcommand{\emult}{\end{multline}}

\def\d{{\rm d}}

\def\b{\mathrm{b}}

\def\lp{{\ell}_{\rm Pl}}

\def\f{\frac}

\def\d{\textrm{d}}

\def\ul{\underline}

\def\rmax{\rho_{\rm max}}

\def\lp{l_{\rm Pl}}

\def\ld{l_{\Delta}}
\def\d{{\rm d}}

\def\svp{\Psi(\nu,\phi)}
\def\ld{\lambda}
\def\b{b}

\allowdisplaybreaks

\usepackage{enumerate}
\begin{document}

\title{Numerical loop quantum cosmology: an overview}
\author{Parampreet Singh}
\email{psingh@phys.lsu.edu}
\affiliation{Department of Physics and Astronomy, Louisiana State University,
Baton Rouge, Louisiana 70803, USA}


\begin{abstract}
A brief review of  various numerical techniques used in loop quantum cosmology and results is presented. These include the way extensive numerical simulations shed insights on the resolution of classical singularities, resulting in the key prediction of the bounce at the Planck scale in different models, and the numerical methods used to analyze the properties of the quantum difference operator and the von Neumann stability issues. Using the quantization of a massless scalar field in an isotropic spacetime as a template, an attempt is made to highlight the complementarity of different methods to gain understanding of the new physics emerging from the quantum theory. Open directions which need to be explored with more refined numerical methods are discussed. 
\end{abstract}

\maketitle

\section{Introduction}
It is generally believed that a quantum theory of gravity will provide important insights on the resolution of singularities, the boundaries of spacetime in Einstein's theory of General Relativity (GR). Since we do not yet have a complete theory of quantum gravity, a pragmatic approach is to use known techniques of quantum gravity to understand quantization of spacetimes with reduced degrees of freedom, such as cosmological models. This provides a test bed not only for methods of full quantum gravity, but potentially opens new avenues to eventually link quantum gravity with phenomenology and observations. An early attempt in this direction, based on Wheeler's ideas on geometrodynamics
was Wheeler-DeWitt quantum cosmology.  However, due to various mathematical and physical hurdles, in particular lack of guidance from a more complete theory, it met with little success on the resolution of singularities. Though Wheeler-DeWitt quantization yielded consistent behavior in the infrared regime (i.e. at the classical scales) in agreement with GR, unfortunately its predictions agreed with GR even in the ultraviolet (large spacetime curvature) regime. As an example, a semi-classical state peaked in a macroscopic universe at late times in Wheeler-DeWitt theory, remains peaked on the classical trajectory throughout the evolution and encounters the big bang singularity.

In recent years, various shortcomings of Wheeler-DeWitt quantum cosmology have been overcome in the framework of loop quantum cosmology (LQC) \cite{as}, which is a quantization of spacetimes with finite degrees of freedom based on loop quantum gravity (LQG), a candidate for non-perturbative canonical quantization of gravity. As in Wheeler-DeWitt theory, LQC is based on Dirac's method of constraint quantization. Instead of geometrodynamics variables in Wheeler-DeWitt theory, the classical gravitational phase space is labelled by
the symmetry reduced versions of Ashtekar-Barbero connection $A^i_a$ and its conjugate triad $E^a_i$, the phase space variables in LQG. Classical Hamiltonian constraint is expressed in terms of the elementary variables of quantization, the holonomies of connection and triads, and is  quantized. An inner product is found using methods of group averaging \cite{ga} which leads to a physical Hilbert space, and physics is extracted using a family of Dirac observables. This algorithm has been carefully carried out for various models in LQC. At the level of quantum Hamiltonian constraint, a striking difference between Wheeler-DeWitt quantum cosmology and LQC turns out to be the following.  Unlike Wheeler-DeWitt theory, where the quantum Hamiltonian constraint yields a {\it differential} equation on a continuum spacetime, the quantum evolution equation in  LQC turns out to be a {\it difference} equation in geometrical variable(s). The origin of discrete quantum evolution operator lies in  the discrete action of holonomy operators, which captures the non-local nature of field strength of the connection on the underlying quantum geometry.

LQC started with the seminal works of Bojowald more than a decade ago, which indicated resolution of singularities, albeit at a kinematical level \cite{mb_old,mb1}. These ideas were put on a firm mathematical footing by Ashtekar, Bojowald and Lewandowski in Ref. \cite{abl}.  A loop quantization of cosmological spacetime, where the inner product, Dirac observables and the resulting physical implications using states in the physical Hilbert space were first made available, was performed for a spatially flat isotropic model with a massless scalar field \cite{aps1,aps2,aps3}. For this model, classical solutions are generically singular, and there is no possibility, as is sometimes true in models with potentials, to fine tune any parameter to avoid the singularity. Thus, it is an ideal arena to test whether methods of LQG can resolve the singularity. The massless scalar field also allows to overcome a conceptual difficulty to understand evolution in constraint quantization. In the absence of external time, scalar field serves the role of internal clock, and physics can be extracted using relational observables.\footnote{In the absence of fields, this role can be played by geometrical degrees of freedom. An example is the case of vacuum Bianchi-I model, where predictions can be extracted by treating one of the scale factors as internal time.}  A detailed analysis of resulting physics through sophisticated analytical and numerical methods, demonstrated that the
big bang singularity is resolved and replaced by a quantum bounce when energy density $(\rho)$ of the scalar field became equal to a critical value, $\rho \approx 0.41 \rho_{\rm{Planck}}$ \cite{aps3}. At small spacetime curvature (or large volume), the quantum Hamiltonian constraint, a difference operator with uniform discretization in volume, approximates the Wheeler-DeWitt quantum constraint, and hence agrees with GR. Robustness of results on genericity of  bounce and maximum energy density were confirmed with an exactly soluble model in LQC \cite{acs}. Thus, quantum geometric effects in LQC, not only cure the singularity problem, but also lead to a physically consistent infrared behavior. In the early quantizations of LQC \cite{mb_old,mb1,abl}, the latter was a serious problem, which was noted in different works \cite{unruh,kvp,ck1}. As we will discuss later, this limitation is tied to the structure of quantum difference equation in the old quantization, which is uniformly discrete in eigenvalues of triads (proportional to area). It turns out to have  difficulties with the von Neumann stability issues at large volumes and leads to a critical density which depends on the phase space variables in such a way that the bounce could occur even at very small spacetime curvature \cite{aps2}.\footnote{There are additional problems with such a critical density which will be discussed in Sec. V.} It turns out that for the isotropic models, the quantization which yields a difference equation with uniform discreteness in volume, which is also referred as the improved quantization, is the only one which results in the correct infrared behavior for various types of matter and a critical density which is a fundamental constant \cite{cs08}.

With the success of improved quantization of LQC \cite{aps3}, quantization of isotropic spacetimes with massless scalar field and positive spatial curvature \cite{apsv,warsaw_closed}, negative spatial curvature \cite{kv,szulc}, and, positive \cite{kp,pa} and negative cosmological constant \cite{bp}
were performed.\footnote{In above works, numerical techniques  to find physical states in the quantum theory were largely developed by Tomasz Pawlowski.} Each of these models introduced non-trivial subtleties in the quantization procedure, and unlike the case of spatially flat model, they could not be solved exactly. Extensive numerical simulations in these models confirmed the robustness of results on bounce,\footnote{In these works, occurrence of bounce was studied by computing expectation values of volume operator. Recently, bounce in massless scalar model has been established by computing quantum probabilities using consistent histories approach \cite{craig_singh1}. A similar analysis for Wheeler-DeWitt theory predicts probability of encountering a singularity as unity \cite{craig_singh2}.} and revealed various new features of rich physics resulting from quantum geometry. Recently, a rigorous quantization of Bianchi-I model with a massless scalar field has been performed \cite{awe2}, which overcomes problems with an earlier quantization \cite{chiou}. Quantization of Bianchi-II \cite{awe3} and Bianchi-IX models \cite{we}  have also been proposed, and developments on cosmological models, have also provided insights on loop quantization of Schwarzschild spacetimes \cite{bh,bh1,bh2,bh3}.
 As in the case of isotropic models, to loop quantize these spacetimes rigorously and consistently requires overcoming several mathematical hurdles. From the perspective of numerical methods, one of the difficulties is that unlike the case of isotropic models, the quantum Hamiltonian constraint in anisotropic and black hole spacetimes turns out to be a partial difference equation. With the non-uniform discretization in different variables, numerical analysis of loop quantum Hamiltonian constraints in these models is far more challenging than the isotropic models.

In the above discussion we focused on the role of numerical methods following the analysis in Refs. \cite{aps1,aps2,aps3}. However, use of numerical techniques to understand properties of quantum difference equation in old quantizations of LQC are dated earlier \cite{bojo_pc,bd1,unruh,kvp,ckb,ck1,time,ck2,date,sv_semi}. In these works, carried out before inner product and physical Hilbert space in LQC became available, the thrust of most of the analysis  was to gain insights on viability of particular quantizations in LQC by using ideas of von Neumann stability to relate solutions of quantum difference equation at large volumes to the Wheeler-DeWitt equation. These works also introduced  notions of pre-classicality of solutions to identify the underlying conditions for a consistent infra-red behavior \cite{bd1,ck1}. It is important to note some important differences in the usage of 
von Neumann stability analysis in LQC, and its conventional treatments in computational physics. Unlike 
the conventional treatments on von Neumann stability issues where the continuum partial differential equation (PDE) is fundamental and the goal is to understand whether a finite difference equation obtained by discretization of PDE provides a good approximation, in LQC it is the quantum difference equation which is fundamental and the goal of stability analysis is to verify whether in the large volume limit (or at small spacetime curvatures),  solutions of quantum difference equation agree with those of the Wheeler-DeWitt equation. Note that the quantum difference equation obtained in LQC is not obtained by a discretization of the Wheeler-DeWitt theory, and the goal of stability analysis is not to compare its solutions with Wheeler-DeWitt equation at ultraviolet (or small volume) scales.  Using these ideas,  
the application of von Neumann stability techniques has provided  useful insights on the infrared behavior of quantizations of various models in LQC. Currently, techniques are being developed to overcome problem of non-uniform discretizations in partial quantum difference equations in anisotropic and black hole spacetimes in LQC \cite{khanna_sabh,nelson2}.

Another avenue where numerical techniques in LQC have been extensively used is in the effective dynamics approach. This is based on the observation that the underlying quantum dynamics in various models, is approximated by an effective spacetime description derived from an effective Hamiltonian \cite{jw,vt}. In this approach, instead of using quantum difference equations, one works with the modified dynamical equations on a continuum spacetime encoding quantum gravity corrections. Thus, one can use standard numerical methods to solve differential equations. Nevertheless, various studies in this approach require a tight control on errors which has been carefully achieved. Examples of such works include studies on dynamics in inflationary and Ekpyrotic potentials \cite{svv,csv}, computation of classical probabilities in inflation \cite{sloan,ck_infl}, assisted and multi-field inflationary dynamics \cite{rs},  Gowdy models \cite{hybrid} and various investigations in anisotropic spacetimes \cite{cv,csv,gs1,cm2,gs2}. Effective dynamics approach has provided important insights on various issues in LQC. These include effects on cosmological perturbations \cite{cmb}, genericity of singularity resolution \cite{ps09,sv,ps11}, and constraints on possible discretizations of the quantum difference operator \cite{cs08,cs09}, which shed light on issues noted in stability analysis of quantum difference equation.

The goal of this manuscript is to provide a brief overview of various numerical methods used in LQC. These range from those yielding physical evolution in the quantum theory to those related to using von Neumann stability ideas on quantum difference equation in LQC. Due to space limitations, it is difficult to go in details of all the diverse models and their developments, and some omissions are unfortunately inevitable. These include  detailed discussions on various results from effective dynamics where conventional numerical methods to solve differential equations suffice, factor ordering and symmetrization issues\footnote{Various results discussed in this manuscript are unaffected by these issues. For readers interested in a detailed discussion on these issues we refer to Refs. \cite{kp,nelson1}.}, and in-depth discussions on pre-classicality issues which were developed before a rigorous formulation of LQC was available.  Some of these results, such as issues of pre-classicality, are covered in detail in a recent review on numerical methods in LQC which provides a complementary view to various developments \cite{numrev}. To make our discussion simpler to follow and to integrate results from different methods, we take the strategy of explaining different methods and results using a single model -- quantization of spatially flat isotropic model with a massless scalar, and provide summary of main results of other models.

The plan of this manuscript is as follows. In Sec. II, we describe the basic setting of loop quantization using the example of a spatially flat Friedmann-Robertson-Walker model with a massless scalar field which was rigorously quantized in LQC in Refs. \cite{aps1,aps2,aps3}. This model serves as an excellent tool to illustrate various details of the quantization procedure and properties of quantum theory, and has served as a template for quantization of models with spatial curvature \cite{apsv,warsaw_closed,kv,szulc}, in presence of cosmological constant \cite{bp,kp,pa} and anisotropic models \cite{awe2,bianchi1_madrid,awe3,we}. After describing the classical phase space in Sec. IIA, we briefly discuss the quantum theory in Sec. IIB and obtain quantum Hamiltonian constraint as a difference equation. We conclude this section with a brief discussion of Wheeler-DeWitt quantization and the classical limit of loop quantum evolution equation. (Readers who are interested only in the numerical techniques can skip Sec. II and refer to eq.(2.18)). Sec. III is devoted to numerical methods used to obtain physical solutions of the quantum theory. We focus on the way numerical simulations are carried out in spatially flat isotropic model with a massless scalar field \cite{aps1,aps2,aps3}, and discuss two methods to obtain physical states: one by using a fast Fourier transform (FFT) and another using evolution in internal time in Sec. IIIA and Sec. IIIB respectively.  We summarize the results of spatially flat isotropic model in Sec. IIIC, and the other models in Sec. IIID. In Sec. IV, we discuss the von-Neumann stability analysis. After explaining various subtleties and differences with the conventional stability analysis, we discuss examples of massless scalar, positive cosmological constant in LQC, and difference equation in the old quantization in LQC \cite{abl,aps2}. The latter case serves to illustrate limitations of isotropic models which do not yield a quantum difference equation which is uniformly discrete in volume, such as the old quantization in LQC \cite{mb1,abl} and lattice refined models which allow arbitrary discretizations \cite{lattice}. Sec. V deals with a brief discussion of effective spacetime description of LQC, and the way effective dynamics sheds insights on issues in stability analysis and uniqueness of discretization in isotropic LQC. We conclude with a summary and outlook in Sec. VI.

\section{Loop quantum cosmology: spatially flat isotropic model}
In this section we provide a brief overview of loop quantization of spatially flat Friedmann-Robertson-Walker (FRW) spacetime with a massless scalar field. Our discussion is based on the analysis in Refs. \cite{aps1,aps2,aps3,acs}.  We start with a discussion of the classical phase space, relation of connection and triad variables with the metric variables, and the classical Hamiltonian constraint.  We then discuss kinematical aspects of the quantum theory and demonstrate the way physical Hilbert space and Dirac observables are obtained. We obtain quantum difference equation which turns out to be uniformly discrete in volume, and discuss its classical limit, which agrees with the Wheeler-DeWitt equation.

\subsection{Classical framework}
The spacetime metric for the spatially flat ($k = 0$) isotropic homogeneous spacetime  is given by
\be
\d s^2 = - N^2 \, \d t^2 + a^2(t) \, d {\bf{x}}^2
\ee
where $N$ is the lapse function and $a(t)$ denotes the scale factor of the universe. The spatial manifold for the $k=0$ model can be non-compact with ${\mathbb{R}^3}$ topology, or compact with a 3-torus (${\mathbb{T}^3}$) topology. If the topology is non-compact, we must introduce a fiducial cell ${\cal V}$ as an infrared regulator to avoid divergences in spatial integrations in order to define a Hamiltonian framework. An obvious but important consistency requirement is that the choice of the cell ${\cal V}$ must not affect the physical predictions of quantities invariant under the change of the cell.\footnote{As will be discussed later, this requirement can be used as one of the criteria in selection of a unique discretization of quantum difference evolution operator in LQC \cite{cs08}.} The fiducial volume of the cell ${\cal V}$ is given by $V_o = \int_{\cal V} \d^3 x \, {\sqrt{\q}}$, where $\q$ denotes the determinant of the fiducial metric $\q_{ab}$ on the spatial manifold defined by the co-moving coordinates $x^a$. The physical metric on the spatial manifold $q_{ab}$ is related to the fiducial metric as $q_{ab} = a^2 \q_{ab}$, and the physical volume of the cell ${\cal V}$ is given by $V = a^3 V_o^{1/3}$.

The gravitational phase space variables in LQG are the Ashtekar-Barbero SU(2) connection $A^i_a$ and its canonical conjugate, the densitized orthonormal triads $E^a_i$, which satisfy
\be
\{A^i_a(x), E_j^b (y)\} \, = \, 8 \pi G \gamma \, \delta^a_b \, \delta^i_j \delta^3(x,y) ~.
\ee
Here $\gamma \approx 0.2375$ is the Barbero-Immirzi parameter whose value is set by the black hole thermodynamics in LQG \cite{meissner}. The connection and the triad variables are related to the conventional metric variables in the following way. The triad $E^a_i$ are related to the spatial metric as $E^a_i E^b_j = q \, q^{ab}$, and the Ashtekar-Barbero connection is related to the extrinsic curvature $K_{ab}$ as $A^i_a = \Gamma^i_a + \gamma K^i_a = \Gamma^i_a + \gamma e^{b i} K_{ab}$, where $\Gamma^i_a$ denotes the spin connection which vanishes for the $k=0$ model, $e^{a i}$ is the undensitized triad, and $K^i_a$ is the extrinsic curvature 1-form.

Given the symmetries of the isotropic and homogeneous FRW spacetime, the connection $A^i_a$ and triad $E^a_i$ can be expressed in terms of an isotropic connection and triad pair $(c, p)$ \cite{abl}:
\be
A^i_a \,= \,c \, V_o^{-1/3} \, \w^i_a , ~~~ E^a_i \,= \,p \, V_o^{-2/3} \, \sqrt{\q} \, \e^a_i ~,
\ee
where $\e^a_i$ denote fiducial triads and $\w^i_a$ are the fiducial co-triads compatible with the fiducial metric $\q_{ab}$. The canonical
pair $(c,p)$ satisfy the Poisson bracket relation $\{c,p\} = 8 \pi G \gamma/3$. Note that the triad can take  positive or negative values depending on the relative orientation of the physical and fiducial triads. Since we will not consider fermions in our analysis, the choice of an orientation represents a gauge freedom which does not affect physical predictions. It will be later fixed by choosing symmetric states in the quantum theory. It is also useful to note the relation between $(c,p)$ and the metric variables. The triad $p$ is related to scale factor as $|p| = V_o^{2/3} a^2$, and {\it only on the classical solutions of GR}, $c = \gamma V_o^{1/3} \dot a/N$, where a `dot' denotes derivative with respect to time $t$.

It turns out that quantum theory is considerable simpler if we express the gravitational phase space in terms of a canonical pair $(\b, \nu)$ related to $(c,p)$ as,
\be\label{bnudef}
\b \, = \, \f{c}{|p|^{1/2}}, ~~~ \nu \, = \, \f{|p|^{3/2}}{2 \pi \, \gamma \, G \hbar} \, \mathrm{sgn}(p)~,
\ee
where $\mathrm{sgn}(p)$ is +1 if physical and fiducial triads have same orientation, and is -1 if the orientation is opposite. The conjugate variables $\b$ and $\nu$ satisfy,
\be
\{\b,\nu\} = \f{2}{\hbar} ~,
\ee
where $\hbar$ in the denominator is an artifact of the definition of $\nu$ in (\ref{bnudef}). It should be noted that though $\nu$ is a measure of physical volume $V = |p|^{3/2}$, it has dimensions of length. The variable $\b$ has dimensions of inverse length, and {\it only in the classical theory} is related to the Hubble rate $H$ of the scale factor as $\b = \gamma H = \gamma \dot a/a$ (for the lapse $N=1$). The matter part of the phase space, for the massless scalar model under consideration,  is labelled by the scalar field $\phi$ and its conjugate momentum $p_\phi$,
which satisfy $\{\phi, p_\phi \} = 1$. The conjugate momentum is related to $\dot \phi$ as $p_\phi = V_o \, a^3 \dot \phi$ (for $N=1$).
The classical phase space is thus four dimensional identified by $(\b, \nu; \phi, p_\phi)$.

Before we proceed to the Hamiltonian constraint, it is useful to note the way these variables transform under freedoms related to the choice of spatial coordinates and fiducial volume $V_o$ which brings forward the advantage of using $\b$ instead of $c$ in the framework.   The first freedom allows a  rescaling of coordinates $x \rightarrow x' = \alpha x$ without affecting the metric. Under this freedom, various variables considered above transform as
\be
a \, \rightarrow \, \alpha^{-1} a, ~~ V_o \, \rightarrow \, \alpha^3 V_o, ~~ V \, \rightarrow \, V, ~~ (c,p) \, \rightarrow \, (c,p), ~~ (\b, \nu) \, \rightarrow \, (\b,\nu) ~~ {\rm{and}} ~~ (\phi, p_\phi) \, \rightarrow \,  (\phi, p_\phi) ~.
\ee
Thus, under the rescaling of coordinates, both $c$ as well as $\b$ are invariant. However, under the freedom of the choice of fiducial cell, ${\cal V} \rightarrow \beta^3 {\cal V}$, one obtains
\be\label{scalingeq}
a \, \rightarrow \, a, ~~ V_o \, \rightarrow \, \beta^3 V_o, ~~ V \, \rightarrow \, \beta^3 V, ~~ (c,p) \, \rightarrow \, (\beta c, \beta^2 p), ~~ (\b, \nu) \, \rightarrow \, (\b,\beta^3 \nu) ~~ {\rm{and}} ~~ (\phi, p_\phi) \, \rightarrow \,  (\phi, \beta^3 p_\phi) ~.
\ee
It is to be noted that under the rescaling of fiducial cell, $\b$ is invariant but $c$ is not.
Further, it is straightforward to check that if we consider a phase space variable $P_m = c |p|^m$,  then it is invariant under rescaling of fiducial cell only when $m = -1/2$, i.e. when $P_m = \b$. The variable $P_m$ is used in the lattice refinement scheme with $-1/2 < m < 0$ \cite{lattice}\footnote{The usage of the term `lattice' here should not be confused with the one in lattice loop quantum cosmology in Ref. \cite{wilson_lattice}, where it is used in the context of a lattice of homogeneous universes to investigate inhomogeneities.}, and this observation will be useful for later discussion (in Sec. V) to gain insights on the consistency of improved quantization \cite{aps3} and inconsistencies of other quantizations in isotropic LQC.

Before we write the classical Hamiltonian constraint in terms of gravitational phase space variables, we note that the  scalar field $\phi$ satisfies $\Box \phi = 0$, and is used as internal time in LQC. It is then natural to choose the lapse $N = a^3$ such that the time variable $\tau$ satisfies $\Box \tau = 0$. Unless noted specifically, in the following, we will work with this choice of lapse function. Using this lapse function, the classical Hamiltonian constraint for the isotropic homogeneous model with a massless scalar field becomes\footnote{The presence of $\hbar$ in the following classical constraint  is an artifact of the definition of $\nu$ in eq.(\ref{bnudef}).}
\be \label{class_cons}
C^{\rm{class}}_H = - 3 \pi G \hbar^2 \b^2 \nu^2 + p_\phi^2 \approx 0 ~.
\ee
Using the classical Hamilton's equations, it is straightforward to obtain the dynamical equation from the above Hamiltonian constraint,
\be
\f{\p \nu}{\p \phi} \, = \, \pm \sqrt{12 \pi G} \, \nu
\ee
whose integration yields the classical trajectories:
\be\label{traj}
\phi = \pm \f{1}{\sqrt{12 \pi G}} \ln \f{\nu}{\nu_c} + \phi_c
\ee
where $\nu_c$ and $\phi_c$ are constants of integration. In the classical theory, we thus obtain expanding and contracting branches. The expanding branch encounters big bang singularity at $\nu = 0$ in the backward evolution at a certain value of internal time $\phi$. Similarly, the contracting branch ends in big crunch singularity in the future evolution.

\subsection{Quantum framework}

In LQG, elementary variables are given by the holonomies of connection $A^i_a$ and fluxes of triad $E^a_i$. In LQC, to define holonomies we consider straight edges $\mu \e^a_k$. The holonomy of a connection $c$ turns out to be,
\be
h_k^{(\mu)} \, = \, \cos \left(\f{\mu c}{2}\right) \mathbb{I} - 2 i \, \sin \left(\f{\mu c}{2}\right) \f{\sigma_k}{2}
\ee
where $\mathbb{I}$ is a unit $2\times2$ matrix and $\sigma_i$ denote Pauli matrices. Similarly, fluxes are computed across the face of the cell ${\cal V}$. Due to homogeneity, the flux integral turns out to be proportional to triads \cite{abl}. Thus, the elementary variables for quantization are the elements of holonomies $N_\mu := e^{i \mu c/2}$ and triads $p$. Since $\mu$ is arbitrary, $N_\mu$ generate an algebra of almost periodic functions of $c$.  The gravitational part of the kinematical Hilbert space ${\cal H}_{\rm{kin}} = {\cal H}_{\rm{kin}}^{\rm{grav}} \otimes {\cal H}_{\rm{kin}}^{\rm{matt}}$ is obtained by finding a representation of holonomy flux algebra using Gel'fand-Naimark-Segal construction, which results in  ${\cal H}_{\rm{kin}}^{\rm{grav}}$ as a space of square integrable functions on the Bohr compactification of the real line: $L^2(\mathbb{R}_{\rm{Bohr}},\d \mu_{\rm{Bohr}})$ \cite{abl}. Unlike the gravitational sector, the matter part of kinematical Hilbert space is obtained by quantizing matter with methods of Fock quantization. In contrast to the Wheeler-DeWitt quantization, where the gravitational part of the kinematical Hilbert space is $L^2(\mathbb{R},\d c)$, normalizable states in
${\cal H}_{\rm{kin}}^{\rm{grav}}$ can be expressed in terms of a countable sum of $N_\mu$, with $\langle N_\mu|N_\mu' \rangle = \delta_{\mu \mu'}$.  It should also be noted that unlike Wheeler-DeWitt theory, there is no operator corresponding to $c$ (or its curvature) in LQC. Instead, its information is contained in the corresponding holonomy operator.  The action of the operators $\hat N_\mu$ on states $\Psi(\mu)$ in the triad representation is translational:
$\hat N_\zeta \, \Psi(\mu) = \Psi(\mu + \zeta)$, where $\zeta$ is a constant. In contrast, the action of $\hat c$  in Wheeler-DeWitt theory is differential in the  triad representation (as in the standard Schr\"odinger mechanics). Thus, even at the kinematical level, there are significant differences between LQC and Wheeler-DeWitt quantizations. The most important difference being that the continuum differential geometry common to GR and Wheeler-DeWitt theory is replaced by a discrete quantum geometry in LQC.

The next step in the quantization procedure is to express the classical Hamiltonian constraint in terms of the elementary variables for quantization. In terms of triads and holonomies, the gravitational part is given by
\be
C_{\rm{grav}} \, = \, \gamma^{-2} \, V_o^{-1/3} \, \epsilon^{i}_{~j k} \, \e^a_i \, \e^b_j \, |p|^2 F_{ab}^{~k} ~.
\ee
Here $F_{ab}^{~k}$ is field strength of the connection, expressed in terms of holonomies over a square plaquette $\Box_{ij}$ with length $\bar \mu V_o^{1/3}$
\be
F_{ab}^{~k} = - 2 \, \lim_{Ar \Box \rightarrow 0} {\rm{Tr}} \left(\f{h_{{\Box}_{ij}} - \mathbb{I}}{Ar\Box} \, \tau^k\right) \, \w^i_a \, \w^j_b ~,
\ee
where $h_{\Box_{ij}} = h_i^{(\bar \mu)} h_j^{(\bar \mu)} (h_i^{\bar \mu})^{-1} (h_j^{\bar \mu})^{-1}$. Before one promotes $C_{\rm{grav}}$ to a quantum operator, it is important to note that the limit $\Box_{ij} \rightarrow 0$ does not exist in the quantum theory. The non-existence of this limit is tied to the underlying quantum discreteness in LQG, and is consistent with the fact that in LQG there exist no operators corresponding to connection or its curvature. 
The underlying quantum discreteness allows the loop $\Box_{ij}$ to be shrunk only to a minimum area, given by the lowest non-zero eigenvalues of the area operator in LQG,  $\Delta \lp$ where $\Delta = 4 \sqrt{3} \pi \gamma$. This constrains the parameter $\bar \mu$ to \cite{awe2}
\be\label{mubar}
\bar \mu^2 = \f{\Delta \lp^2}{|p|} ~.
\ee
The dependence of $\bar \mu$ on triad $p$ makes the  action of resulting field strength operator,
\be
\hat F_{ab}^{~k} \Psi(\mu) \, = \, \epsilon_{ij}^{~k} V_o^{-2/3} \, \w^i_a \w^j_b \widehat{\left(\f{\sin^2(\bar \mu c)}{\bar \mu^2}\right)} \, \Psi(\mu) ~,
\ee
less straightforward than the simple translation action of elements $\exp(i \sigma c/2)$ on the eigenstates of the triad. It turns out that the action simplifies in the volume representation $(\nu)$. In the eigenbasis of $\hat \nu$ operator,  the action of elementary variables $\hat V$ and its conjugate  $\widehat{\exp(i \lambda \b)}$, where $\lambda$ is a parameter with dimensions of length, is given by
\be
\hat V \, |\nu\rangle \, = \, 2 \pi \gamma \lp^2 \, |\nu| \, |\nu\rangle, ~~~ \widehat{\exp(i \lambda \b)} \, |\nu\rangle \, = \, |\nu - 2 \lambda \rangle ~.
 \ee
The parameter $\lambda$ is related to the area gap as $\lambda^2 = \Delta \lp^2$. It thus directly captures the discreteness of the underlying quantum geometry.

The resulting action of the gravitational part of the quantum Hamiltonian constraint yields a  difference equation in uniform steps of volume. For the case of a massless scalar field, the action of total Hamiltonian constraint operator $\hat C_H = \hat C_{\rm{grav}} + 16 \pi G \hat C_{\rm{matt}}$ is given by,
\be \label{cons1}
\partial_\phi^2 \, \Psi(\nu,\phi) \, = \, 3 \pi G \, \nu \, \f{\sin \lambda \b}{\lambda} \nu \, \f{\sin \lambda \b}{\lambda} \, \Psi(\nu,\phi) \, =: \, - \, \Theta \Psi(\nu,\phi) \\
\ee
where $\Theta$ is a positive definite,  internal time independent difference operator, with the following action:
\be\label{thetacons}
\Theta \Psi(\nu, \phi) \, := \, - \f{3 \pi G}{4 \lambda^2} \, \nu \left((\nu + 2 \lambda) \Psi(\nu + 4 \lambda) - 2 \nu \Psi(\nu,\phi) + (\nu - 2 \lambda) \Psi(\nu - 4 \lambda)\right) ~.
\ee
The physical states $\Psi(\nu,\phi)$ are chosen to be
symmetric under the change in orientation of triads, i.e. $\Psi(\nu,\phi) = \Psi(-\nu,\phi)$. In the absence of fermions, as is the case in this analysis, this symmetry requirement eliminates a large gauge freedom induced by the parity operator $\hat \Pi$ with the action $\hat \Pi \Psi(\nu) = \Psi(-\nu)$,  and associated with the choice of the triad orientation. Note that the form of the quantum constraint is very similar to as in the Klein-Gordon theory, where $\phi$ plays the role of time and $\Theta$ acts like a spatial Laplacian operator, and as in the Klein-Gordon theory, physical states can be decomposed into positive and negative frequency subspaces.

In order to extract physics, we need to find an inner product. It can be obtained by using group averaging procedure \cite{ga}, or demanding that the action of Dirac observables, $\hat V|_\phi$, the volume at internal time $\phi$, and $\hat p_\phi$ be self-adjoint. The physical inner
product turns out to be
\be
\langle \Psi_1| \Psi_2\rangle \, = \, \sum_\nu \, \bar \Psi_1(\nu,\phi_o) |\nu|^{-1} \Psi(\nu_2,\phi_o) ~.
\ee
The action of Dirac observables on states $\Psi(\nu,\phi)$ is given by
\be
\hat V|_{\phi_{o}} \Psi(\nu,\phi) \, =  \, 2 \pi \gamma \lp^2 \, e^{i \sqrt{\Theta} (\phi - \phi_o)} |\nu| \Psi(\nu,\phi_o), ~~ {\rm{and}} ~~ \hat p_\phi \Psi(\nu,\phi) \, = \, - i \hbar \, {\p_\phi} \, \Psi(\nu,\phi) = \sqrt{\Theta} \Psi(\nu,\phi) ~.
\ee
The action of the volume observable $\hat V|_{\phi_o}$ can be understood as considering  the state at $\phi = \phi_o$, multiplying by volume $V$, and evolving the state to $\phi$. Dirac observables preserve the positive and negative frequency subspaces, and
thus it suffices to consider positive frequency states as the physical states, which satisfy $- i \hbar \, \partial_\phi \Psi(\nu, \phi) \, = \, \sqrt{\Theta} \Psi(\nu,\phi).$  

This provides us the physical Hilbert space ${\cal H}_{\rm{phys}}$ which is the space of normalized positive frequency states $\Psi(\nu,\phi)$ which are symmetric under the change in orientation of triads.
Since $\Theta$ is a difference operator, physical states have support on lattice $\nu = \pm \epsilon + 4 n \lambda$, where $\epsilon \in [0,4)$, and the Hilbert space is decomposed into subspaces ${\cal H}_\epsilon$ which are preserved by the evolution. To extract physics we can restrict ourselves to any of the sectors $\epsilon$ and the results turn out to be independent of this choice.  From the point of view of the big bang singularity, the most interesting sector is $\epsilon = 0$. (Numerical results on singularity resolution discussed in next section, correspond to this sector).
With the availability of a self-adjoint quantum evolution operator, Dirac observables, and the inner product, we have a rigorous framework to extract physics from this quantum theory.\\

The exercise carried out above can be repeated in a straightforward way for the Wheeler-DeWitt theory \cite{aps2,aps3,acs}. One can find the kinematical Hilbert space, action of quantum Hamiltonian constraint and Dirac observables $\hat V|_\phi$ and $\hat p_\phi$, and the inner product. A crucial difference is that unlike in LQC, the underlying geometry in Wheeler-DeWitt theory is not discrete and
the quantum constraint (analogous to (\ref{cons1}), turns out to be \cite{aps3,acs}
\be\label{wdweq}
\p_\phi^2 \Psi(\nu, \phi) \, = - \underline{\Theta} \, \Psi(\nu,\phi) := \, 12 \pi G \, \nu \, \p_\nu \nu \, \p_\nu \, \Psi(\nu, \phi) ~.
\ee
Thus, unlike LQC, where the evolution equation is a difference equation with uniform steps in volume, in Wheeler-DeWitt theory the evolution equation turns out to be  a differential equation.


We now show that the LQC quantum difference equation leads to the Wheeler-DeWitt equation in the large volume limit. For smooth wavefunctions, at $\nu \gg \lp$, we can expand $\Psi(\nu \pm 2 \lambda, \phi)$ in eq. (\ref{thetacons}) as,
\be
\Psi(\nu \pm 2 \lambda, \phi) \, = \, \Psi(\nu,\phi) \, \pm \, 4 \lambda \, \p_\nu \Psi(\nu,\phi) \, + \, \f{1}{2} \, (4 \lambda)^2 \, \p_\nu^2 \Psi(\nu,\phi) \, \pm \f{1}{6} \, (4 \lambda)^3 \, \p_\nu^3 \Psi(\nu,\phi) + ...
\ee
Substituting in eq.(\ref{cons1}), we obtain,
\be
\p_\phi^2 \, \Psi(\nu,\phi) \, \simeq \, 12 \pi G \, \nu \left(\p_{\nu} \Psi(\nu,\phi) \, + \, \nu \p_{\nu}^2 \Psi(\nu,\phi)\right) + O\left(\lambda^3 \p_\nu^3 \Psi(\nu,\phi)\right) ~.
\ee
If the wavefunction is slowly varying, in the approximation that the terms of the order $\lambda^3 \p_\nu^3 \Psi$ can be neglected, we obtain eq.(\ref{wdweq}).
Thus, in the limit of large volume (small spacetime curvature), the quantum Hamiltonian constraint in LQC which is a quantum difference equation, yields the Wheeler-DeWitt equation.


\section{Numerical techniques for obtaining physical states}
In the previous section, we showed the way loop quantization of a spatially flat isotropic model results in a quantum Hamiltonian constraint which is a difference equation with
uniform discreteness in volume. Here we discuss numerical methods to determine physical states satisfying the quantum Hamiltonian constraint \cite{aps2,aps3}. We first describe a method based on calculation of symmetric eigenfunctions and computing a fast Fourier transform (FFT) to determine physical states. This is followed by description of another method, involving specification of an initial state peaked on a classical trajectory at late times in a macroscopic universe and evolved in internal time $\phi$ using quantum evolution equation (\ref{cons1}). We summarize results from these methods, and for other models in Sec. IIIC and Sec. IIID respectively.

\subsection{Physical states by FFT method}
This method is based on obtaining physical states  by noting that any symmetric state in the physical Hilbert space can be expressed in terms of symmetric eigenfunctions $e_k^{(s)}$ of the $\hat \Theta$ operator:
\be\label{integral}
\Psi(\nu,\phi) \, = \, \int_{-\infty}^\infty \, \d k \, \tilde \Psi(k) \, e_k^{(s)} \, e^{i \omega \phi} ~,
\ee
where $\tilde \Psi(k)$ is a suitable function to define state profile, such as a Gaussian function.
By numerically finding symmetric eigenfunctions, one can perform the FFT in the above equation and obtain the physical state. Since this method relies on the properties of eigenfunctions, we discuss some of the features of the latter below.
Symmetric eigenfunctions are obtained by considering symmetric combinations of eigenfunctions of the $\Theta$ operator: $\Theta e_\omega(\nu) = \omega^2 e_\omega (\nu)$, where $\omega^2 = 12 \pi G k^2$. The eigenfunctions $e_\omega(\nu)$ have a 2-fold degeneracy on the Hilbert space ${\cal H}_{|\varepsilon|}^{\rm{grav}}$ (or on ${\cal H}_{-|\varepsilon|}^{\rm{grav}}$), a subspace of physical Hilbert space with states having support on the lattice  ${\cal L}_{|\varepsilon|}$ (or ${\cal L}_{-|\varepsilon|}$). The way symmetric eigenfunctions are constructed depends on the choice of the sector ($\varepsilon$). For $\varepsilon = 0 \, {\rm or} \, 2$, lattices ${\cal L}_{|\varepsilon|}$ and ${\cal L}_{-|\varepsilon|}$ coincide with each other under the transformation: $\nu \rightarrow - \nu$, and symmetric eigenfunctions can be constructed from either of the lattices. Whereas, for $\varepsilon \neq 0, 2$, these lattices are distinct, and hence symmetric eigenfunctions are constructed using eigenfunctions $e_\omega(\nu)$ on positive as well negative lattices. In this case,  eigenfunctions have a 4-fold degeneracy and symmetric eigenfunctions are constructed by a linear combination of $e_{\pm |k|}$ on positive and negative lattices.

\begin{figure}[tbh!]
\includegraphics[angle=270,width=0.8\textwidth]{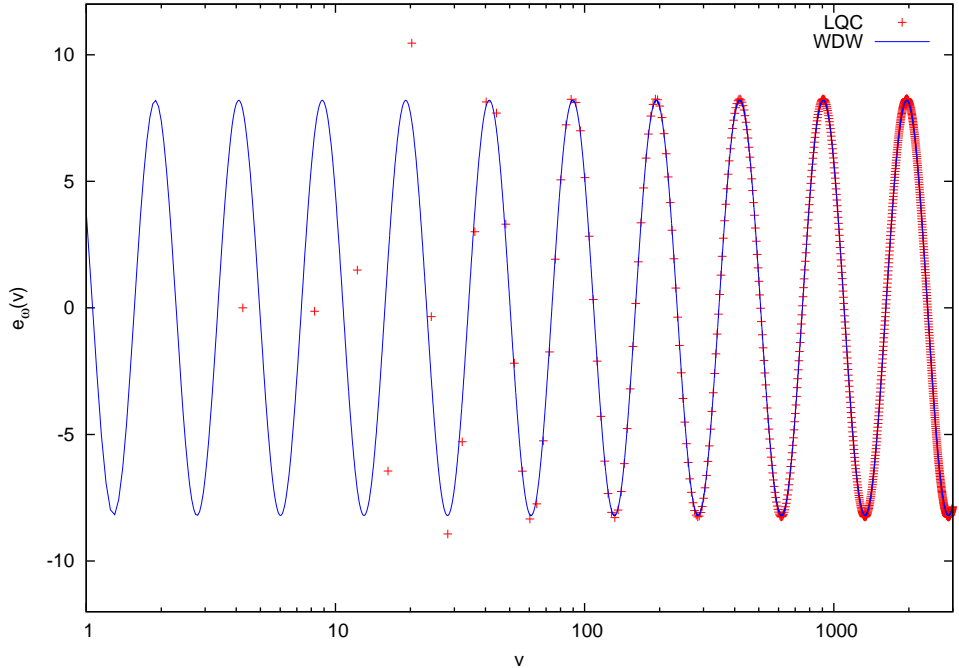}
\caption{An eigenfunction of quantum Hamiltonian constraint in LQC is compared with that of the Wheeler-DeWitt theory (solid line), for $\omega = 50$. Here $v$ corresponds to $\nu/\lambda$ in Planck units. Eigenfunctions approach each other at large volumes. At small volumes, eigenfunction in LQC decays exponentially.}
\end{figure}

To obtain these symmetric eigenfunctions, we first compute eigenfunctions $e_\omega(\nu)$ numerically for a given $\varepsilon$. Since at large volumes $(V \gg \lp^3)$, $\Theta \rightarrow {\ul{\Theta}}$,  eigenfunctions $e_\omega(\nu)$ are expected to approach the Wheeler-DeWitt eigenfunctions ${\ul{e}}_\omega(\nu)$. This indeed turns out to be true. In Fig. 1, we show the behavior of an eigenfunction in LQC and compare it with that of the Wheeler-DeWitt theory. Here we note that each eigenfunction ${\ul{e}}_\omega(\nu)$ of the Wheeler-DeWitt equation (\ref{wdweq}) can be expressed in a linear combination of basis functions ${\ul{e}}_{|k|}(\nu)$ as,
\be\label{eigenlargev}
{\ul{e}}_\omega(\nu) \, = \, A \, {\ul{e}}_{|k|} (\nu) + B \, {\ul{e}}_{-|k|} (\nu) ~,
\ee
where $A$ and $B$ are constant coefficients,
and at large volumes $e_\omega(\nu) \simeq  {\ul{e}}_\omega(\nu)  + O(\lambda^2/\nu^2)$.
It is interesting to note that the eigenfunctions $e_\omega(\nu)$ decay exponentially at small volumes. The volume at which exponential decay occurs is numerically found to be proportional to the value of $\omega$ (or the choice of $p_\phi$). Recently, the exponential damping of eigenfunctions in the quantum regime has been studied analytically, which provides additional insights on the behavior of eigenfunctions \cite{craig}. It turns out that the volume at which exponential decay occurs corresponds to the volume at the bounce of the universe in LQC.

After obtaining symmetric eigenfunctions, we need to choose a function $\tilde \Psi(k)$ in the integral (\ref{integral}). We are interested in those states which at late times (or large volumes) correspond to a universe with a small spacetime curvature described by GR. We  consider a sharply peaked state in volume as well as its conjugate $\b$ at late times on a classical trajectory. Due to relation between $\b$, $p_\phi$ and $V$ implied by the classical constraint (\ref{class_cons}), it suffices to choose a state which is peaked in volume and $p_\phi$. To be concrete let us choose such a state at a large value of $p_\phi$ in the expanding branch: $|p_\phi^*| \gg \sqrt{G} \hbar$. If $k^*$ (which is negative in the expanding branch) corresponds to the value of $p_\phi$, via $p_\phi = - \sqrt{12 \pi G} \hbar k$, at which the state is peaked  then for a small dispersion $\sigma$ we choose
\be
\tilde \Psi(k) \, = \, e^{-(k - k^*)^2/2 \sigma^2} \, e^{-i \omega \phi^*} ~.
\ee
The volume at which state is peaked at time $\phi = \phi_o$ is determined by the value of $\phi^*$ on using the equation for classical trajectory (\ref{traj}). To compute $\Psi(\nu,\phi)$, one finds symmetric eigenfunctions in a range, say $k^*-10\sigma, k^*+10 \sigma$, by discretizing the interval in a large number of points (approximately $2^{12}$). The physical state  $\Psi(\nu,\phi)$ is computed by a FFT, and, the norm $\norm \Psi^2$ and expectation values of Dirac observables, $\hat V|_\phi$ and $\hat p_\phi$ can be evaluated. We discuss the resulting physics in Sec. IIIC.

\begin{figure}[tbh!]
\includegraphics[angle=270,width=0.45\textwidth]{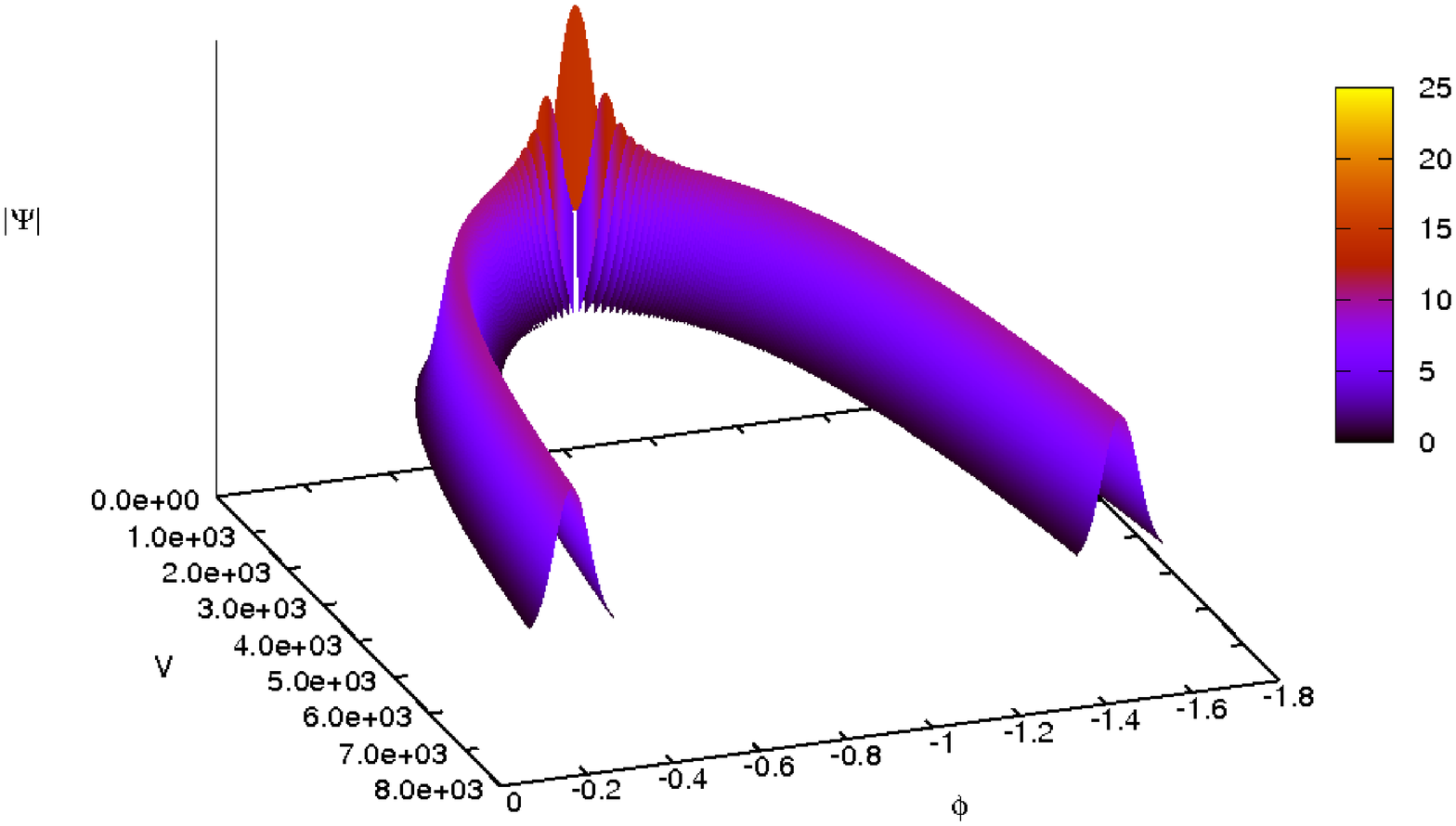}
\includegraphics[angle=270,width=0.45\textwidth]{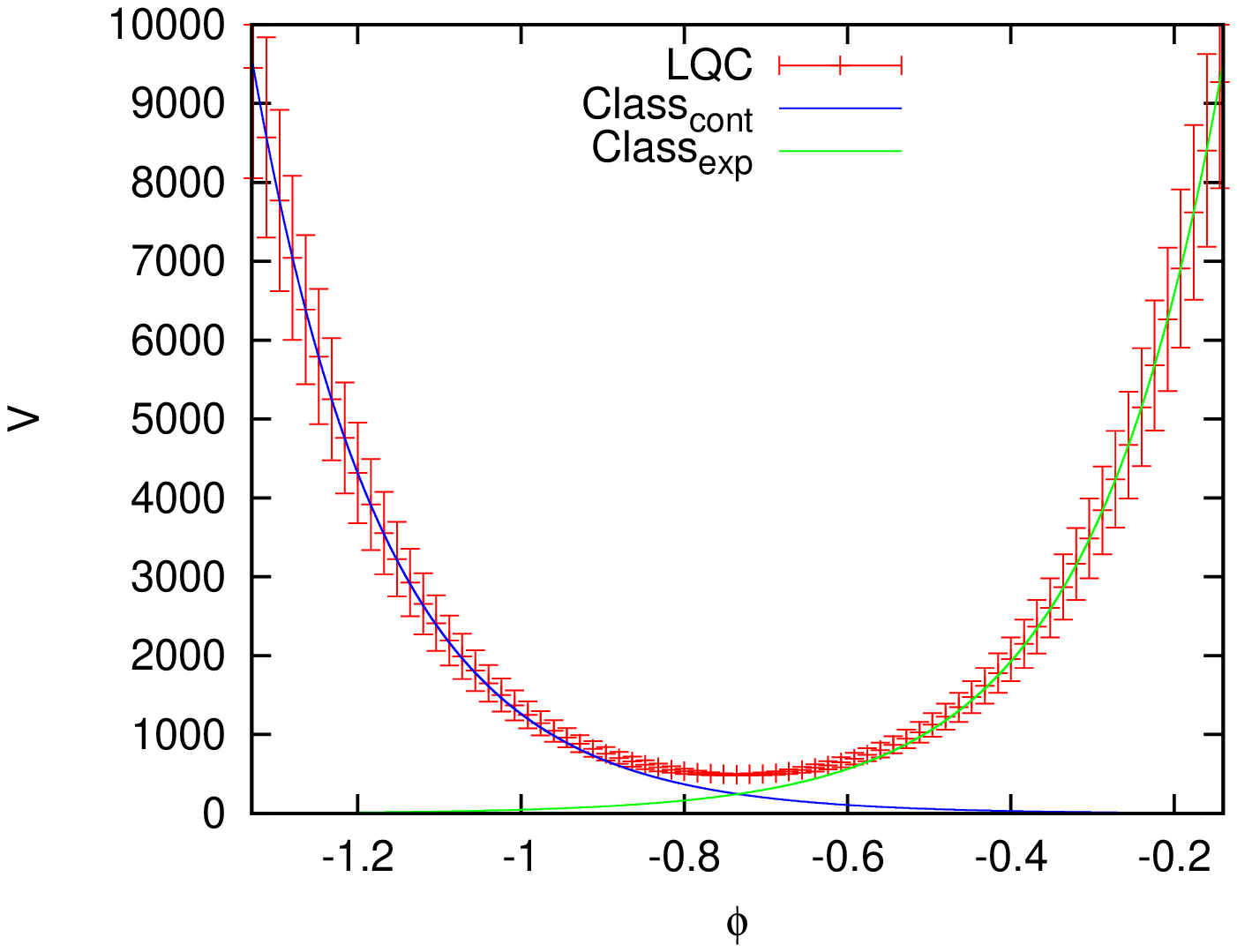}
\caption{The plot  shows a section of the evolution of the amplitude of a physical state $\Psi(\nu,\phi)$ obtained using internal time evolution described in Sec. IIIB, and the resulting expectation values of the volume observable. The initial state is a solution of the Wheeler-DeWitt equation with a Gaussian profile peaked at $v^* = 22400$ and $p_\phi^* = 1500$ (in Planck units), and is evolved using difference equation in LQC. Unlike the result of the state evolved in Wheeler-DeWitt theory (shown in Fig. 3), the state in LQC does not encounter singularity, but bounces when energy density becomes equal to $\rmax \approx 0.41 \rho_{\rm{Planck}}$. The bounce in this simulation occurs at $v_b \approx 488$, where $v_b$ denotes value of $v$ at which bounce occurs. At small spacetime curvature, expectation values of volume observable are peaked on disjoint classical trajectories shown by solid lines. }
\end{figure}

\subsection{Physical states by evolution in internal time}
This method deals with the specification of an initial state peaked at a large values of $p_\phi$ and volume at $\phi = \phi_o$ with suitable boundary conditions, and evolve it in the internal time $\phi$ using the quantum difference equation (\ref{cons1}).\footnote{For a detailed discussion of analytical issues involved in this construction, we refer the reader to Refs. \cite{aps2,aps3,madhavan}.}  In this approach, a numerical computation of symmetric eigenfunctions is not required. Rather one has to numerically solve a large number of coupled equations in $\phi$, in a finite domain of numerical integration. To specify boundary conditions, one considers the quantum constraint equation in the form $ i \p_\phi \Psi(\nu,\phi) = \sqrt{\Theta} \Psi(\nu,\phi)$, which at large volumes is approximated by $\p_\phi \Psi(\nu,\phi) = \sqrt{12 \pi G} \nu \p_\nu \Psi(\nu,\phi)$. Using a discretized version of this equation, boundary conditions are specified permitting only
outgoing solutions. In all the numerical simulations carried out in Refs. \cite{aps2,aps3}, the boundary was chosen such that the value of the wavefunction at the boundary was negligible compared to its peak value. 

In numerical simulations of spatially flat isotropic model, three methods have been used to specify
initial data $\Psi|_{\phi_o}$ and $\p_\phi \Psi|_{\phi_o}$ at $\phi = \phi_o$. These are \cite{aps2,aps3}:
\begin{enumerate}
\item A Gaussian state normalized with respect to inner product in the Wheeler-DeWitt theory, which is peaked in $p_\phi$ and volume for a macroscopic universe. The initial state is a minimum uncertainty state in $V$ and $\b$.
\item A solution of Wheeler-DeWitt equation is chosen as an initial state with Gaussian profile peaked at a classical trajectory at small spacetime curvature. It is chosen to minimize uncertainty in $\phi, p_\phi$.
\item A similar method as the previous one with a variation taking into account the fact that eigenfunctions of $\hat \Theta$ in large volume limit
satisfy eq.(\ref{eigenlargev}). To obtain this behavior, initial state is constructed by a phase rotation of eigenfunctions of the Wheeler-DeWitt theory. The phase required for a suitable rotation is determined numerically.
\end{enumerate}

\begin{figure}[tbh!]
\includegraphics[angle=270,width=0.5\textwidth]{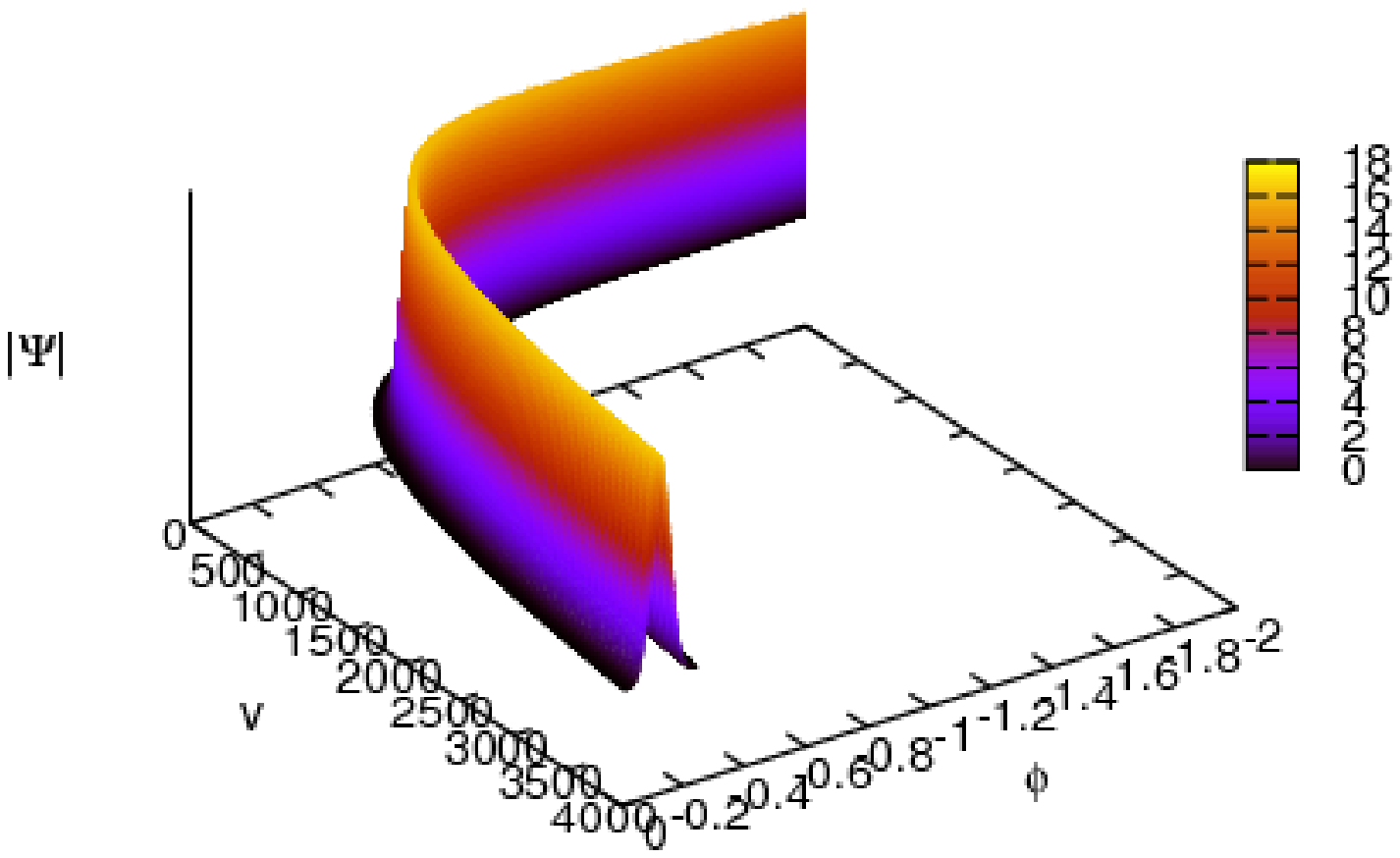}
\includegraphics[angle=270,width=0.45\textwidth]{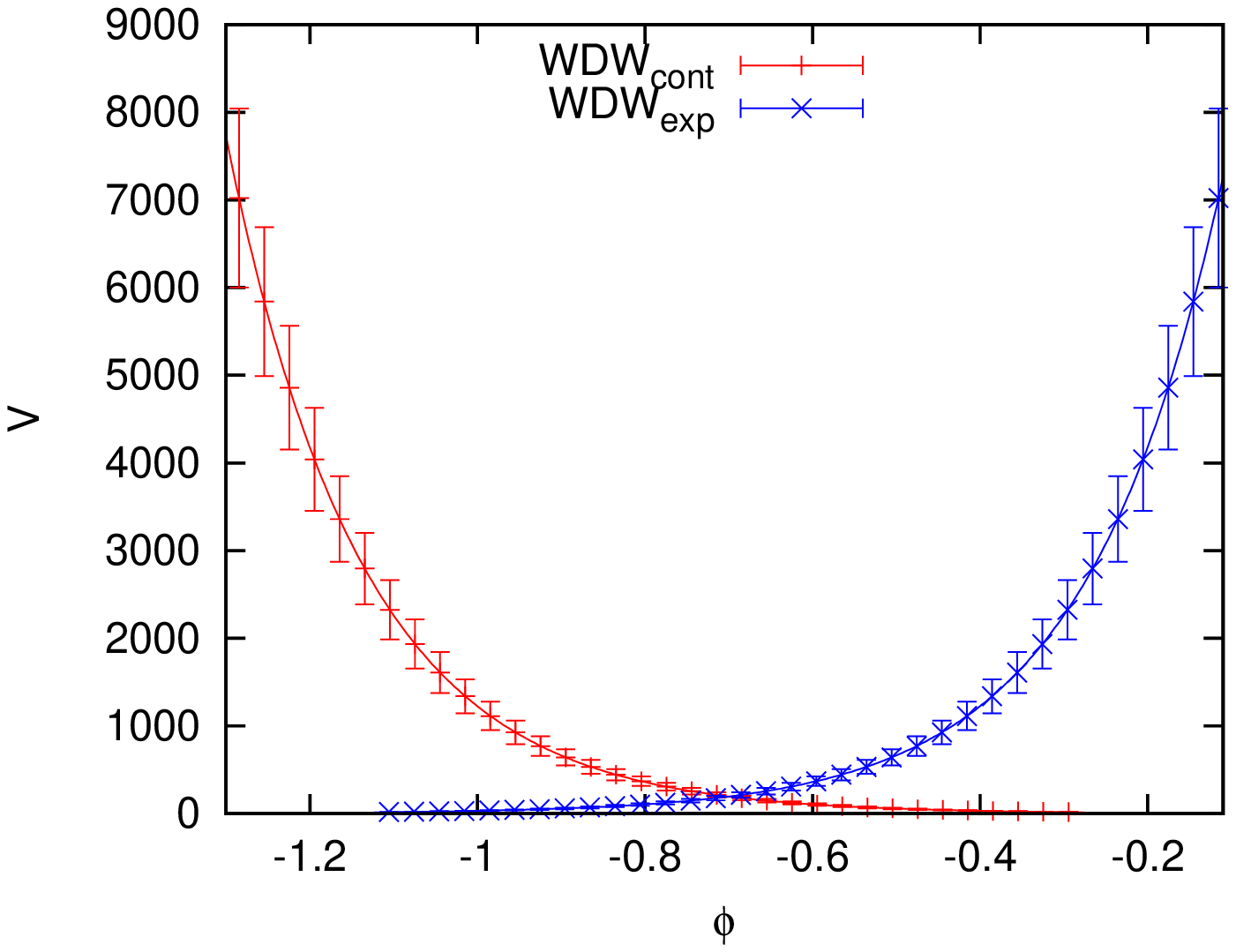}
\caption{The evolution of initial state chosen in Fig. 2 is shown in Wheeler-DeWitt theory for comparison. The state peaked on the classical expanding trajectory evolves to the big bang singularity and does not bounce. The expectation values of volume observable are also shown for two different simulations, one corresponding to expanding branch and the other to the contracting branch. The states in expanding and contracting branches in Wheeler-DeWitt theory are peaked on classical trajectories (shown by solid lines), which are singular and disjoint. }
\end{figure}

With the above initial data, numerical simulations were performed for a large range of $p_\phi$ ranging from 100--10000 (in Planck units), for varying dispersions ($\sim$ 1$\%$ -- 10$\%$)  in Refs. \cite{aps2,aps3}. More recently, these simulations have been carried out for a much smaller values of $p_\phi$ ($\sim$ 5 -- 20), and  higher dispersions ($\sim$ 15$\%$ -- 30$\%$) \cite{dgs}. In the numerical simulations performed in Refs. \cite{aps2,aps3}, numerical errors in internal time discretization were controlled by refining `time' step such that
\be
\norm{\Psi_1 - \Psi_2} \leq \norm \Psi_2 (\Delta \phi) \epsilon
\ee
where $\epsilon$ is a very small number, and, $\Psi_1$ and $\Psi_2$ correspond to wavefunctions computed with two discretization of $\phi$: $\Delta \phi$ and $\Delta \phi/2$. In simulations carried out in Ref. \cite{dgs}, consistency of the  discrete time step has been found by taking into account Courant-Friedrichs-Lewy (CFL) condition for given properties of the initial state.
In various simulations in different works, a typical evolution between $\phi = 0$ to $\phi = -2$ is typically performed by discretizing time interval in $10^5$--$10^6$ steps.\\

We conclude the discussion of Sec. IIIA and Sec. IIIB with a brief comment on the two methods to obtain physical states. Though both methods can be equivalently used to obtain physical states, they may have some advantages and disadvantages depending on the model considered.
As we see above, unlike the method to find physical states by FFT discussed in Sec. IIIA, one needs to carefully address the issues of domain of dependence, convergence and stability in obtaining physical states by the method evolution. Thus, in comparison to the FFT method where physical states are obtained once one has obtained eigenfunctions, this method may  appear more involved. However, a  notable advantage of evolution method lies in the fact that it is computationally economical since it has less memory requirements in comparison to the FFT method. A naive implementation of the FFT method for spacetimes with more degrees of freedom, such as Bianchi-I model, can raise the computational cost significantly. On the other hand, the latter method, at least for some simple models, is comparatively straightforward to be parallelized.

\subsection{Key features of results and their robustness}
We summarize results from the numerical methods described above to obtain physical states in spatially flat isotropic model with a massless scalar. 
We also discuss, in brief, results from the 
spatially curved models with massless scalar \cite{apsv,kv,szulc}, with positive and negative cosmological constants \cite{pa,bp} and from preliminary works in the presence of anisotropies.

\begin{itemize}
\item The main result obtained using numerical simulations based on FFT and state evolution methods, is that physical states in LQC do not encounter classical big bang singularity. Rather, expectation values of the volume observable, $\langle \hat V|_\phi \rangle$ reveal that universe bounces from a forward evolution in contracting branch to an expanding branch (or vice-versa) when the energy density of the scalar field becomes $\rho \approx 0.41 \rho_{\rm{Planck}}$. 
The bounce occurs independent of the choice of $p_\phi$ or the sector $\epsilon$. This behavior can be seen in Fig. 2 where amplitude of the wavefunction in LQC is plotted versus $v = \nu/\lambda$ and $\phi$ and expectation values of volume observable are compared with classical trajectories. The initial state is  peaked at $v^* = 22400$ (where $v = \nu/\lambda$) with $p_\phi^* = 1500$ (in Planck units), and evolved using method to consider initial state as a solution of Wheeler-DeWitt equation in Sec. IIIB. The lattice in this simulation is chosen to be $\varepsilon = 0$. To
compare with the Wheeler-DeWitt theory, we show the evolution of the same initial state specified in the expanding branch using Wheeler-DeWitt equation (\ref{wdweq}) in Fig. 3. Unlike the state in LQC which bounces, one finds that the state in Wheeler-DeWitt theory is peaked on the classical trajectory at all times and encounters the big bang singularity at $v = 0$ in the  past evolution. Note that unlike, Wheeler-DeWitt theory where expectation values agree with classical solution at all times, in LQC expectation values show a non-singular bounce which joins two disjoint trajectories of classical GR.

\item Classical GR turns out to be a  good approximation to loop quantum evolution till the energy density reaches a percent of the Planck value. Significant deviations from classical trajectory occur when energy density becomes larger than this value. In regime where energy density is smaller than above value, a physical state in LQC is peaked on a classical trajectory in the expanding branch after the bounce, and on a classical contracting branch in pre-bounce regime,  corresponding to the same value of $p_\phi$. Quantum gravitational effects bridge two disjoint classical solutions. Thus, LQC yields GR in the infrared limit and new Planck scale physics devoid of singularity in the ultraviolet regime. This behavior can be seen from plot of expectation values  in Fig. 2, which shows that at large volumes, there is an excellent agreement between LQC and classical GR, where as at small volumes there are significant departures.

\item Physical states remain sharply peaked through out the evolution, as is also clear from Fig. 2. Depending on the method used and properties of initial state, there is only a little variation in behavior of relative fluctuations across the bounce. Subsequent to the analysis in Refs. \cite{aps3,acs}, strong constraints on change in relative fluctuation of volume and $p_\phi$ observables across the bounce have been obtained analytically for states which are similar as well as more general than those considered in numerical simulations \cite{cs_recall,kp2,cm1}. These analytical results confirm the numerical observation that if one considers an initial state peaked in a classical macroscopic universe at late times, it evolves via LQC quantum constraint to a state across the bounce with similar features at early times.
\item Norm of the state is preserved and expectation value of $\hat p_\phi$ observable remains constant throughout the bounce.
\item The expectation value of volume observable at the bounce coincides with the value at which eigenfunctions become exponentially damped \cite{aps2,aps3}. This result has also been confirmed analytically \cite{craig}.

\item Extensive numerical simulations confirm that underlying quantum dynamics in LQC can be approximated by an effective spacetime dynamics even in the Planck regime for states which peak in a macroscopic universe at late times.  This behavior is shown in Fig. 4 and is discussed in detail in Sec. V.
\end{itemize}

\noindent
{\bf Remark 1:} Spatially flat isotropic model with a massless scalar field can be solved exactly in the $\b$ representation \cite{acs}. Results obtained analytically confirm with those from the numerical simulations in the $\nu$ representation performed earlier. In particular, using exactly soluble model one finds that the bounce occurs for generic states in the physical Hilbert space, {\it not necessarily semi-classical ones}, in the sense that the expectation values  $\langle \hat V|_\phi \rangle$ are bounded below. The energy density of the scalar field has a finite upper bound given by $\rmax = 3/(8 \pi G \gamma^2 \lambda^2) \approx 0.41 \rho_{\rm{Planck}}$, which agrees with the value of critical energy density at which bounce occurs in all numerical simulations.\\

\noindent
{\bf Remark 2:} In the old quantization of LQC \cite{bh1,abl}, the quantum difference equation turns out to be uniformly
discrete in the eigenvalues of triads. This occurs if while relating area of the loop over which holonomies are considered to the minimum eigenvalue of area operator, $\bar \mu$ is considered as a constant $(\mu_o)$. The resulting action of the elements of the holonomies $\exp(i \mu_o c)$ is by uniform translations on triad eigenvectors, a feature shared by the action of the quantum Hamiltonian constraint in the old quantization.
A rigorous quantization of this model was performed in Ref. \cite{aps1,aps2}. As in the improved quantization, there is a bounce, but with an important difference. Unlike the bounce discussed above, in the old quantization of LQC, bounce does not occur at a universal value of energy density. For the massless scalar model, the energy density at the bounce turns out to be inversely proportional to $p_\phi$. Also, recall that $p_\phi$ is not invariant under rescaling of fiducial cell (eq.(\ref{scalingeq})).\footnote{As discussed in Sec. V, the infrared problem occurs in this quantization irrespective of problem of rescaling of the fiducial cell.}  These features lead to a serious infrared problem in this quantization. The same problem plagues loop quantizations of isotropic models which do not yield quantum difference equation with uniform discreteness in volume (such as lattice refinement scheme \cite{lattice}). We discuss these issues further in Sec. V.

\subsection{Other models}
In this part, we briefly summarize various numerical results obtained in spatially curved model, spatially flat model with positive and negative cosmological constant, and in presence of inflationary potential, and Bianchi-I model. \\

\noindent
{\bf Spatially curved models:} Following the quantization procedure outlined in Sec. II and numerical techniques discussed above, a rigorous analysis of spatially closed ($k=1$) isotropic model with a massless scalar field was performed in Refs. \cite{apsv,warsaw_closed}. Unlike the $k=0$ model, a classical $k=1$ filled with massless scalar inevitably collapses in future evolution of the expanding branch and encounters a big crunch singularity in a finite time. Thus, there is a physical singularity both in the past and the future evolution. LQC resolves both of these singularities for arbitrary choices of parameters. A technical difference in comparison to the $k=0$ model is that the corresponding quantum difference operator, $\Theta_{k=1}$ (which is uniformly discrete in volume)  has discrete eigenvalues, due to which numerical procedure is more involved. Results from the numerical simulations with states peaked at classical trajectories at late times and evolved with the quantum difference equation show bounce at $\rho \approx 0.41 \rho_{\rm{Planck}}$ before big bang and big crunch are reached. In this model, loop quantum dynamics exhibits a cyclic evolution. Notably, states remain sharply peaked throughout the evolution for many cycles \cite{apsv}. 
As in the case of spatially flat model, an excellent agreement of loop quantum dynamics and effective dynamics obtained from an effective Hamiltonian was found \cite{apsv}.

   Loop quantization of $k=1$ model overcame various limitations of an earlier quantization of the same model in LQC \cite{mb1}, which led to a quantum difference equation with uniform discreteness in triad eigenvalues. These limitations were first noted in the work of Green and Unruh \cite{unruh}, who found that eigenfunctions of the quantum constraint operator do not exhibit consistent infrared properties. In particular, they do not decay at large volumes where recollapse of the classical $k=1$ universe is expected. These problems were resolved in the
quantization proposed in Ref. \cite{apsv,warsaw_closed}. We will later discuss the way this issue is related to von Neumann stability of the quantum difference equation in the old and improved quantizations in LQC.

The strategy for loop quantization of $k=0$ and $k=1$ models runs into some technical difficulties for $k=-1$ model \cite{kv,szulc}. The existing quantization of this model considers holonomies of extrinsic curvature rather than the connection, and the resulting quantum difference operator is not essentially self-adjoint. Nevertheless, physical states have been studied by computing the eigenfunctions and taking a FFT \cite{kv}. These states reveal existence of bounce when energy density becomes equal to $\rmax \approx 0.41 \rho_{\rm{Planck}}$. The quantization also shares various nice features of $k=0$ and $k=1$ models, including states remaining sharply peaked at all values of internal time, and consistent ultraviolet and infrared behavior. Effective dynamics also provides a consistent picture of the evolution of physical states in LQC. \\

\noindent
{\bf Cosmological constant:} Spatially flat model with a massless scalar field has been loop quantized with a positive \cite{pa} and a negative cosmological constant \cite{bp}. As in the spatially flat case with a massless scalar, the difference equation turns out to be uniformly discrete in volume. For the  case of negative cosmological constant, classical evolution predicts a recollapse of the universe at large volume, similar to the case of $k=1$ model, and apart from past big bang, there is a big crunch singularity in the future evolution. The quantum difference operator turns out be self-adjoint, with discrete eigenvalues and its spectrum is numerically found to be non-degenerate. Numerical simulations reveal existence of bounce when total energy density of scalar field and cosmological constant become equal to $\rho \approx 0.41 \rho_{\rm{Planck}}$, and the big bang and big crunch singularities of the classical theory are avoided. Numerical studies with a large range of cosmological constant confirm that the recollapse of the universe occurs at the classically predicted volume. Further, states are found to be sharply peaked throughout the evolution, and effective dynamics is found to be in excellent agreement with quantum evolution \cite{bp}.

A rigorous quantization of spatially flat model with massless scalar and positive cosmological constant reveals various interesting features of the quantum difference operator \cite{kp,pa}. In the classical theory, for any choice of positive cosmological constant, universe expands to infinite volume. With the massless scalar field playing the role of internal time, the infinite volume is reached in a finite value of internal time, $\phi = \phi_f$. Its one consequence is that the quantum difference operator $\Theta_{\Lambda}$ is not essentially self-adjoint. However, by choosing a self-adjoint extension, quantum evolution can be continued beyond $\phi_f$. Extensive numerical simulations show that the detailed physics depends weakly on the choice of self-adjoint extension. The quantum difference operator $\Theta_\Lambda$ has discrete eigenvalues, however they depend on the choice of self-adjoint extension. Numerical evolution of states reveal that big bang singularity is resolved in the past evolution, and the evolution continues beyond $\phi_f$ where a cycle of contraction starts. Bounces of the universe in different cycles occur when total energy density $\rho \approx 0.41 \rho_{\rm{Planck}}$. The quantum evolution from $\phi = -\infty$ to $\phi = \infty$ is thus composed of infinite cycles of expansion and contraction. Further, numerical simulations find an agreement between loop quantum evolution and effective dynamics. It is interesting to note that a non-trivial physical Hilbert space exists only when value of cosmological constant is smaller than a critical value of $\Lambda = 3/\gamma^2 \lambda^2$ \cite{kp}. Note that in the presence of massless scalar field, if $\Lambda$ becomes equal to critical value, energy density would become greater than $\rcr \approx 0.41 \rho_{\rm{Planck}}$. This observation is also tied to the von Neumann stability of the quantum difference equation, which we discuss in Sec. IV.\\

\noindent
{\bf Inflationary potential:} Quantization of $\phi^2$ inflationary potential in spatially flat isotropic LQC is based on common features with that of the case of positive cosmological constant discussed above \cite{aps4}. In this model, scalar field $\phi$ is not monotonic, however it can be treated as an internal clock locally. At each instant $\phi$, the quantum difference operator $\Theta$ is equivalent to the one in the case of positive cosmological constant, and is hence not essentially self-adjoint. For evolution, one thus has to carefully choose self-adjoint extensions between different `time' slices. Preliminary numerical analysis with states which lead to a macroscopic universe at late times with this model indicates existence of bounce at $\rho \approx 0.41 \rho_{\rm{Planck}}$, and validity of the effective Hamiltonian approach till the scale of bounce.\\

\noindent
{\bf Bianchi-I spacetimes:} Recently, a loop quantization of Bianchi-I model with a massless scalar field has been rigorously performed by Ashtekar and Wilson-Ewing \cite{awe2}. Due to three spatial triads, the structure of non-singular quantum difference operator is much richer in comparison to the isotropic model. In this model, it is a partial difference equation in geometrical variables. Due to the underlying complexity of the partial difference equation, numerical evolution of physical states in this model are yet to be performed. One of the technical hurdles arises due to non-uniformity of discrete variables, which we discuss in Sec. IV.  

Numerical simulations have been performed for vacuum Bianchi-I model with a different quantization proposed earlier by Chiou \cite{chiou}. Though this quantization suffers from fiducial cell scaling and infrared problems \cite{cs09}, it nevertheless serves as a good playground to develop and test numerical methods beyond isotropic description. Detailed properties of the quantum theory, such as essential self-adjointness of the quantum difference operator, were found and techniques to obtain physical solution were introduced in Ref. \cite{madrid_bianchi1}. In this analysis, one of the geometric variables plays the role of internal time, and relational dynamics between different geometric variables shows absence of singularity and existence of bounce. States remained sharply peaked throughout the evolution.  It is to be noted that dynamics obtained from effective Hamiltonian constraint is shown to be a good approximation to the quantum evolution.

\section{Von-Neumann stability analysis of difference equations}

A useful way to understand  various properties of the quantum difference operator $\hat\Theta$ in LQC at large volumes, and its relation to the corresponding operator $\ul{\hat\Theta}$ in the Wheeler-DeWitt theory is by
using ideas of von Neumann stability analysis. In various branches of physics, this analysis plays an important role in obtaining faithful numerical solutions of partial differential equations (PDE's) by discretization. Using von Neumann analysis, it is possible to determine whether a particular discretization of a PDE is stable, which by Lax-Richtmyer equivalence theorem is a necessary and sufficient condition for a ${\it consistent}$ finite difference scheme to be {\it convergent}. Here consistency and convergence are used in the conventional sense of numerical methods (see for eg. \cite{book}). By consistency we imply that a smooth solution of a PDE is an approximate solution of finite difference equation obtained by its discretization when discreteness parameters go to zero, and convergence implies that a solution of finite difference equation approximates a solution of the PDE.

In a conventional von Neumann analysis, for a given discretization of a PDE, say 1+1 dimension in space and time, one proceeds by a Fourier decomposition of the solution of finite difference equation in space which results in a recursion relation in time. Using this algebraic equation one can determine the amplification factor $g$ between the values of the solution at two neighboring time steps. Von Neumann analysis requires that $|g| \leq 1$ for all the roots of the algebraic equation. There are some important differences in the conventional usage of von Neumann stability analysis in various branches of physics and in LQC. These are noted below:
\begin{enumerate}
\item Unlike in the conventional applications of von Neumann analysis where differential equation is fundamental, in LQC it is the difference equation which is fundamental. The discretization in geometric variable(s) is fixed by the underlying quantum geometry. Thus, in order to numerically solve the quantum Hamiltonian constraint in LQC, there is no freedom in changing the Planck scale discretization of quantum geometry to enhance numerical efficiency.\footnote{Note that there are two types of discreteness studied here in LQC. Apart from the discreteness in quantum geometry resulting from the properties of area and volume operators in LQG, the other discreteness deals with structure of difference equation -- whether it uniformly discrete in volume, area or any other geometric variable. (For a discussion on similar issues in LQG and broader context, we refer the reader to Ref. \cite{carlo}). A quantum difference equation, irrespective of whether it is uniformly discrete in volume or not, shares the same kinematical properties of discrete quantum geometry identified by the eigenvalues of the geometrical operators.} As emphasized in Sec. I, in the von-Neumann stability analysis in LQC, our goal is to compare solutions of difference equations which are uniformly discrete in different geometrical variables obtained from loop quantizations with the Wheeler-DeWitt theory at large volumes to understand the infrared behavior of quantum difference equation in LQC.   It is to be noted that as in conventional numerical methods, there is freedom to choose the discretization in (internal) time labelled by $\phi$ appropriately.\footnote{This freedom is expected to be absent or severely limited if the scalar field is also polymer quantized using methods of LQG, instead of being Fock quantized.} 

\item In the von Neumann analysis, stability issues are analyzed in the limit of discretization parameters going to zero. Since the discretization parameter(s) of geometry in LQC are fixed, one uses von Neumann analysis in the regime where the discreteness scale can be ignored. As an example, for the isotropic model with massless scalar, one can consider a large volume limit $\nu \gg \lp$ or $\nu \rightarrow \infty$. In early works in LQC, different types of limits were used to analyze stability. Above criteria was first used in the work of Cartin and Khanna \cite{ck1}.

\item  There is a difference on the roles of space and time in conventional usage of von Neumann analysis and in LQC. As discussed above, conventionally one performs a Fourier transform in space and analyzes the temporal behavior of amplification factor $g$. In isotropic LQC with massless scalar field as internal time, the Fourier transform is performed in $\phi$, and spatial behavior of amplification factor at large volumes is analyzed. Even for other models so far studied in LQC, such as Bianchi spacetimes and Schwarzschild interior, one identifies one of the geometric variables as `time', and analyzes the behavior of amplification factor in another geometric variable (see for eg. \cite{jung-khanna,ckb2,nelson3}). 

\item Strictly speaking, von Neumann analysis is applicable only for the case of constant coefficients. For variable coefficients, as we will encounter in LQC, one can apply this method using frozen coefficient approximation. In conventional von Neumann analysis, this implies that one considers stability  in a small neighborhood where coefficients in the PDE can be treated as constants up to a leading order. Extra care is needed to interpret results on stability in such a case.
\end{enumerate}

In LQC, the first work incorporating ideas of von Neumann analysis was by Bojowald and Date to understand properties of difference equations resulting from loop quantization \cite{bd1}. At the stage of that early work, knowledge of the inner product and physical Hilbert space, which plays a vital role in ruling out unphysical solutions was not available. Rather, consistency of difference equation with respect to the Wheeler-DeWitt equation was analyzed, to rule out inconsistent quantizations. This analysis gave primary importance to the differential equation, as in the conventional usage of von Neumann analysis, and consistency of difference equation was verified in the limit of discretization parameter becoming zero. This provided one notion of `pre-classicality' in LQC. In a series of works, Cartin and Khanna investigated issues of `pre-classicality' without taking the discreteness parameter to zero, but instead taking an appropriate limit to small spacetime curvature (or large values of volume) \cite{ck1,ck2,ck3}. Solutions having pre-classical behavior have a desirable property expected from solutions of a consistent quantum theory with correct infrared behavior. Such solutions do not display unphysical small scale oscillations in regimes far from the Planck regime. 

Below, we focus on the von Neumann stability issues in the context of spatially flat isotropic models for improved quantization in LQC \cite{aps3}, and compare it with the old quantization in LQC \cite{mb1,abl,aps2}. 
This provides a flavor of the way this technique has been used in LQC to analyze consistency of quantum difference equations \cite{ckb2,nelson3,tanaka}. Our discussion has some parallels with the discussion of isotropic closed model in Ref. \cite{numrev}, and on properties of eigenfunctions in presence of cosmological constant in Ref. \cite{tanaka}, where problems with infrared behavior have been discussed. After discussion of von Neumann stability issues for the spatially flat isotropic model, we summarize main results for other spacetimes in LQC.

\subsection{Stability analysis for flat isotropic model}
Results on the stability analysis for the improved quantization of spatially flat isotropic model with a massless scalar field considered in Sec. II were first reported by Cartin and Khanna \cite{ck_proc}. Recently a comparison of improved and old quantizations in the isotropic model  in perspective of behavior of solutions at large volumes has been made in the presence of cosmological constant \cite{tanaka}, which can be understood using von Neumann stability.  In the following we show the way stability analysis is performed in different models of LQC and contrast between improved and old quantization schemes. As we  discuss below, important lessons and pitfalls of using this analysis can be understood by considering different matter. We first discuss the stability of quantum difference constraint as discussed in Sec. II for a massless scalar field and a positive cosmological constant. We then show that though the quantum difference equation for old quantization is stable for massless scalar, instabilities develop on inclusion of positive cosmological constant. As emphasized before, we note that in LQC it is the quantum difference equation which is fundamental, and stability or instability used below should not be confused with the conventional usage of terms in von Neumann analysis where PDE is fundamental and the goal is to determine whether a discretization of the PDE is stable or unstable.

\subsubsection{Isotropic model with massless scalar}
The  loop quantization of massless scalar field in spatially flat isotropic model was discussed in Sec. II. Let us begin with rewriting of the quantum difference equation (\ref{cons1}) as follows,
\be \label{diff2}
\p_\phi^2 \svp = C_+(\nu) \Psi(\nu + 4 \ld, \phi) + C_0(\nu) \svp + C_-(\nu) \Psi(\nu - 4 \ld)
\ee
where coefficients $C_+, C_0$ and $C_-$ are given by
\be\label{coeff}
C_+(\nu) = \f{3 \pi G}{4 \ld^2} \, \nu (\nu + 2 \ld), ~~ C_-(\nu) = \f{3 \pi G}{4 \ld^2} \, \nu (\nu - 2 \ld), ~~ C_0(\nu) = - \f{3 \pi G}{ \ld^2} \, \nu ~.
\ee
Our goal is to understand the way solution of the above equation changes (or amplifies) at neighboring steps at large volumes. To find this amplification factor, an equivalent short-cut to Fourier transform procedure used in von Neumann analysis is to consider an ansatz (in the large volume limit)
\be\label{ansatz}
\Psi(\nu,\phi) = g^{n/4 \ld} \, e^{i \omega \phi}
\ee
where $n$ is an integer label of the discrete steps, and $g$ is the amplification factor: $\Psi(\nu + 4 \ld)/\Psi(\nu) = g$. Substitution of (\ref{ansatz}) in (\ref{diff2}) yields the following quadratic equation in the amplification factor:
\be \label{quad1}
C_{+}(\nu) g^2 + (C_0(\nu) - \omega^2) g + C_-(\nu) = 0 ~.
\ee
Since we are interested in the large volume limit, we consider the behavior of coefficients in a small neighborhood at large volume, where they are approximated as 
\be \label{coeff1}
C_+(\nu) \, \simeq \, C_-(\nu) \, = \, \f{3 \pi G}{4 \ld^2} \, \nu^2 \, + \, O\left(\f{\nu}{\ld}\right) ~,
\ee
and
\be \label{coeff2}
C_0 - \omega^2 \simeq - \f{3 \pi G}{2 \ld^2} \, \nu^2 + O(1) ~.
\ee
Von Neumann analysis requires both the roots of the quadratic equation (\ref{quad1}) to have modulus bounded by unity.
Substituting (\ref{coeff1}) and (\ref{coeff2}) in eq.(\ref{quad1}), it is straightforward to see that the two roots for the amplitude $g$ are equal to unity. Thus, by von Neumann analysis, the difference equation (\ref{cons1}) is stable at large volumes.
The boundedness of the roots of the quadratic equation (\ref{quad1}), suggests that there will be no sharp change in the solution at large scales. Recall that at large volumes, quantum difference equation for this model leads to the Wheeler-DeWitt equation (\ref{wdweq}). Hence, according to von Neumann stability analysis, eigenfunctions of LQC at large volume should approximate those of Wheeler-DeWitt theory. This behavior is confirmed by Fig. 1 in Sec. III, where we find that eigenfunctions of LQC smoothly approach to those of Wheeler-DeWitt theory.

\subsubsection{Isotropic model: massless scalar with positive cosmological constant}
A rigorous quantization of positive cosmological constant in LQC has been performed recently by Pawlowski and Ashtekar, and extensive numerical studies of the resulting difference equation have been performed \cite{pa} (see also Ref. \cite{kp} for detailed properties of quantum difference operator).  This case is interesting due to various physical and technical reasons, one of it being that since massless scalar has an equation of state $-1$, and that of cosmological constant is $+1$, the matter Hamiltonian spans the whole range of equation of state for matter satisfying weak energy condition. Thus, this model provides a formidable robustness test of the success of the quantization procedure in LQC.
We now discuss the stability issue for this model on the lines of massless scalar field case discussed above. (A similar analysis of properties of solutions of quantum constraint operator with a different choice of lapse, has been performed by Tanaka et al \cite{tanaka}).

In the case of a positive cosmological constant with a massless scalar field, the quantum Hamiltonian constraint takes the form of the following difference equation,
\be \label{diff2}
\p_\phi^2 \svp = C_+(\nu) \Psi(\nu + 4 \ld, \phi) + (C_0(\nu) + \pi \gamma^2 G \Lambda \nu^2)\svp + C_-(\nu) \Psi(\nu - 4 \ld) ~,
\ee
where $C_+, C_0$ and $C_-$ are given by eq.(\ref{coeff}), and are slightly different from the analysis in Ref. \cite{pa} due to factor ordering. Using the ansatz (\ref{ansatz}) in large volume limit, and following the steps outlined in Sec. IVA1, we obtain a similar quadratic equation as (\ref{quad1}), with the only change in the coefficient of the linear term in amplification factor:
\be\label{eqcoeff3}
C_0(\nu) + \pi \gamma^2 G \Lambda \nu^2 - \omega^2 \simeq - \f{3 \pi G \nu^2}{2 \ld^2} \, \left(1 - 2\,\f{\Lambda}{\Lambda_{\rm{crit}}}\right) + O(1), ~~~ {\rm{where}} ~~~ \Lambda_{\rm{crit}} = \f{3}{\gamma^2 \lambda^2} ~.
\ee
Using above equation and (\ref{coeff1}) in the quadratic equation for $g$, we obtain the following roots:
\be
g \, = \, \left(1 - 2 \f{\Lambda}{\Lambda_{\rm{crit}}}\right) \, \pm \left(\left(1 - 2 \f{\Lambda}{\Lambda_{\rm{crit}}}\right)^2 - 1\right)^{1/2} ~.
\ee
For $\Lambda = 0$, the roots of the equation are given by the massless scalar case discussed above. For $\Lambda = \Lambda_{\rm{crit}}$, the amplification factor gives two roots, equal to -1. For $0 < \Lambda < \Lambda_{\rm{crit}}$, both roots are oscillatory with magnitude equal to unity. However, for $\Lambda > \Lambda_{\rm{crit}}$, one of the roots has magnitude greater than unity. Hence von Neumann analysis implies that for $\Lambda > \Lambda_{\rm{crit}}$, eq. (\ref{diff2}) can not be regarded stable at large volumes. 

This observation is interesting. As discussed in Sec. IIIC, it has been shown that for a loop quantization of  massless scalar field with cosmological constant, a non-trivial physical Hilbert space exists only for $\Lambda < \Lambda_{\rm{crit}}$ \cite{kp,pa}. The physical Hilbert space is zero dimensional for $\Lambda \geq \Lambda_{\rm{crit}}$. The slight discrepancy on the viability of the case $\Lambda = \Lambda_{\rm{crit}}$ between the analysis of Ref. \cite{kp,pa} and von Neumann analysis above, results  due to the approximation involved in eq.(\ref{eqcoeff3}), which essentially ignores the role of massless scalar in the vanishing contribution of $\omega^2$ term. If this order of contribution is included, von Neumann analysis  leads to the same constraints on $\Lambda$ as obtained by mathematical analysis of physical Hilbert space in Ref. \cite{kp,pa}. Further, the value of $\Lambda_{\rm{crit}}$ obtained from above stability analysis is the maximum allowed value of cosmological constant consistent with  the maximum allowed energy density ($\rho_{\rm{max}} = 0.41 \rho_{\rm{Planck}}$) in massless scalar field case for a dense subspace of the physical Hilbert space in LQC. In the numerical simulations of semi-classical states for massless scalar with positive cosmological constant, bounce of the scale factor has also been found to always occur at above value \cite{pa}. It will be interesting to explore deeper relation between above observations originating from different directions in this model.

\subsubsection{Old quantization: massless scalar and cosmological constant}
We now discuss the von Neumann stability issue for the quantum difference equation which originates in the old quantization of LQC. In this quantization of LQC, the parameter $\bar \mu$ in eq.(\ref{mubar}) was assumed to be a constant: $\bar \mu = \mu_o$. With this assumption, the resulting quantum difference equation turns out to be uniform in eigenvalues of triad operator (which are proportional to area), rather than the volume operator \cite{abl}. A careful analysis of the eigenfunctions of this quantization revealed problems with the infrared behavior for the case of cosmological constant \cite{kvp}. Problems with the large volume behavior were also reflected in an early quantization of $k=1$ model in LQC by Green and Unruh \cite{unruh}. As discussed in Sec. IIIC, a rigorous quantization with a massless scalar field in spatially flat model was performed in Ref. \cite{aps2}, which demonstrated a bounce. However, critical density turned out to be proportional to inverse of $p_\phi$, and thus, was not invariant under rescaling of fiducial cell. Insights on the problems of this quantization using effective dynamics are discussed in Sec. V.

In the following we first do analysis with a massless scalar, and then include a positive cosmological constant. Unlike the case of the improved quantization in isotropic LQC, results in the presence and absence of positive cosmological constant turn out to be starkly different. (For this analysis, we consider lapse function to be unity, and we work in the factor ordering of Ref. \cite{aps2} with the same value of discreteness parameter $\mu_o = 3 \sqrt{3}/2$).

In the presence of massless scalar field, loop quantization with $\bar \mu = \mu_o$ yield the following quantum difference equation \cite{aps2}
\be\label{diff3}
8 \pi G \,  B (p) \, \p_\phi^2 \, \Psi(\mu,\phi) \, = \, \left(f_+(\mu) \, \Psi(\mu + 4 \mu_o,\phi) \, + \, f_0(\mu) \, \Psi(\mu,\phi) \, + \, f_-(\mu) \, \Psi(\mu - 4 \mu_o) \right)
\ee
where
\be \label{coeff-f}
f_+(\mu) \, = \, \f{1}{2} \, \sqrt{\f{8 \pi}{6}} \f{\lp}{(\gamma \mu_o)^{3/2}} \, \left| |\mu + 3 \mu_o|^{3/2} - |\mu + \mu_o|^{3/2}\right|, ~~ f_-(\mu) = f_+(\mu - 4 \mu_o) ~~ {\rm{and}} ~~ f_0(\mu) = - f_+(\mu) - f_-(\mu)
\ee
and $B(p)$ denotes eigenvalue of inverse volume operator $\widehat{1/|p|^{3/2}}$, given by
\be
B(p) = \left(\f{6}{8 \pi \gamma \lp^2}\right)^{3/2} \left(\f{2}{3 \mu_o}\right)^6 \left(|\mu + \mu_o|^{3/4} - |\mu - \mu_o|^{3/4}\right)^6 ~.
\ee
We consider the regime $\mu$ very large compared to $\mu_o$, and using the ansatz $\Psi(\mu,\phi) = g^{n/4 \mu_o} e^{i \omega \phi}$ in von Neumann analysis, we obtain the quadratic equation
\be \label{quad3}
f_{+}(\mu) g^2 + (f_0(\mu) - 8 \pi G B(p) \hbar^2 \omega^2) g + f_-(\mu) = 0 ~.
\ee
The coefficients of this quadratic equation can be approximated for large $\mu$ as,
\be \label{coeff3}
f_+(\mu) \simeq f_-(\nu) = \f{3}{2} \sqrt{\f{8 \pi}{6}} \f{\lp}{\gamma^{3/2} \mu_o^2} \, \mu^{1/2} + O(\mu^{-1/2})
\ee
and
\be \label{coeff4}
f_0 - 8 \pi G B(p) \omega^2 \simeq - 3  \sqrt{\f{8 \pi}{6}} \f{\lp}{\gamma^{3/2} \mu_o^2} \, \mu^{1/2} + O(\mu^{-1/2}) ~.
\ee
Substituting (\ref{coeff3}) and (\ref{coeff4}) in eq.(\ref{quad3}), one finds that both the roots of the difference equation (\ref{diff3}) are equal to unity. Thus, for the case of massless scalar field,  old quantization appears to yield a similar result for  stability analysis at large volumes as the improved quantization. Since the domain of validity of von Neumann analysis is only for the large volumes, it will be a mistake to extend this conclusion  to small volumes. As we noted above, and will further discuss in the next section,  though the infrared behavior as analyzed from the perspective of von Neumann analysis is similar for two quantizations, old quantization in LQC for massless scalar field leads to unphysical quantum gravitational effects which are easily ruled out.

We now turn to the interesting case of cosmological constant in the old quantization of LQC. Since this case is known to be problematic in the large volume regime \cite{kvp}, we expect von Neumann analysis to yield amplification factor with modulus greater than unity. In the presence
of positive cosmological constant, the quantum difference equation turns out to be
\be\label{diff3}
8 \pi G \,  B (p) \, \p_\phi^2 \, \Psi(\mu,\phi) \, = \, f_+(\mu) \, \Psi(\mu + 4 \mu_o,\phi) \, + \, \left(f_0(\mu) + 2  \left(\f{8 \pi \gamma \lp^2}{6}\right)^{3/2} \Lambda \mu^{3/2}\right) \, \Psi(\mu,\phi) \, + \, f_-(\mu) \, \Psi(\mu - 4 \mu_o)
\ee
where $f_+, f_0$ and $f_-$ are given by eqs.(\ref{coeff-f}). As before, the von Neumann stability analysis yields a quadratic equation for amplification factor, which is
\be \label{quad4}
f_{+}(\mu) g^2 + \left(f_0 - 8 \pi G B(p)\, \hbar^2 \omega^2 + 2  \left(\f{8 \pi \gamma \lp^2}{6}\right)^{3/2} \Lambda \mu^{3/2}\right) g + f_-(\mu) = 0 ~.
\ee
In the regime where $\mu$ is large, $f_+$ and $f_-$ are approximated by eq.(\ref{coeff3}). The coefficient of the linear term of the above quadratic equation approximates to,
\be
f_0 - 8 \pi G B(p) \omega^2   + 2\left(\f{8 \pi \gamma \lp^2}{6}\right)^{3/2} \Lambda \mu^{3/2}\simeq 2 \Lambda  \left(\f{8 \pi \gamma \lp^2}{6}\right)^{3/2} \mu^{3/2} + O(\mu^{-1/2}) ~.
\ee
Using these expressions in (\ref{quad4}), within the order of approximation of large $\mu$, we obtain following two roots
\be
g_1 = 0, ~~~ g_2 = - \f{16 \pi}{9} \Lambda \,  \lp^2 (\gamma \mu_o)^3 \, \mu ~.
\ee
Since $\gamma$ and $\mu_o$ are fixed, for any given value of $\Lambda$, there exists a sufficiently large value $\mu$, such that  $|g_2| > 1$. Thus, this model fails the von Neumann stability test at large volumes. 
This exercise gives a valuable lesson that one must be careful in claiming stability of a quantization scheme, and it is important to understand behavior of the difference equation with different matter. The above limitation of old quantization has been understood in literature in various ways, and is reflection of the bad infrared behavior of the quantization, and is thus ruled out.

We conclude this subsection with two remarks, one on problems with arbitrary discretizations  which are sometimes proposed in loop quantization \cite{lattice}, and other on the closed models. \\

\noindent
{\bf Remark 1:} In literature, arbitrary discretizations have also been considered in lattice refined models inspired by LQC.\footnote{On similar lines, ad-hoc modifications to the Hamiltonian constraint and their effect on bounce have also been investigated \cite{laguna}. Though such studies may be useful to understand the way properties of difference equations are affected by certain terms, being not derived from loop quantization, such studies give little insights on the robustness of bounce in LQC.} For the isotropic models,
these are based on considering holonomies of a general variable introduced in Sec. IIA, $P_m = c |p|^m$ with $-1/2 < m < 0$ \cite{lattice}. It is straightforward to repeat the above analysis for any value of $m$ in the above range for cosmological constant, and one encounters similar difficulties as noted above for the old quantization in LQC. In fact, in isotropic LQC, it is only for the case of $m = -1/2$, which corresponds to $P_m = \b$, that the von-Neumann stability analysis gives positive result for the case of positive cosmological constant. Thus, the discretization in quantum theory is highly restrictive. This observation is in agreement with the one on uniqueness of discretization in LQC obtained in Ref. \cite{cs08} using effective dynamics (discussed in Sec. V). Recall that in Sec. IIA, we discussed that for $m \neq -1/2$, one does not obtain invariance of phase space variable $P_m$. We will discuss in the next section, how these different observations tie up with each other.\\

\noindent
{\bf Remark 2:} As mentioned earlier, based on an early quantization of isotropic models in LQC, Green and Unruh investigated properties of
quantum difference equation and the behavior of eigenfunctions in spatially closed $k=1$ model with a massless scalar field\cite{unruh}. In this model, classical GR predicts a recollapse of the universe at a volume determined by the initial energy density of the scalar field. If the quantum theory leads to a consistent classical behavior at small spacetime curvature, one expects wavefunctions to exponentially decay near the volume at which recollapse occurs. Authors of the above work, did not find this behavior. Instead, they found that in general, solutions grow exponentially at large volumes. It is to be noted that at that stage of analysis, a rigorous quantum theory of spatially closed model in LQC was lacking. In particular, little was known about the inner product which could provide insights on physical and unphysical solutions. Nevertheless, Green and Unruh's work brought to light limitations with the earlier quantization in LQC. As discussed in Sec. III, these challenges were overcome successfully in an improved quantization which demonstrated correct ultraviolet and infrared behavior \cite{apsv}. Extensive numerical simulations showed that a spatially closed LQC universe, does indeed recollapse at the volume predicted by the classical theory, and resolves the singularity via a non-singular bounce. From the
perspective of discretization, it is to be noted that the quantization Green and Unruh used is uniformly discrete in triad eigenvalues, whereas the correct quantization leads to a difference equation which is uniformly discrete in volume. Recently, a von Neumann stability analysis of quantization considered by Green and Unruh shows that one of amplification factors has modulus greater than unity \cite{numrev}. In contrast, both of the roots in quadratic equation for amplification factors in the quantization of Ref. \cite{apsv}, have modulus equal to unity, as in the flat isotropic model.

\subsection{Anisotropic spacetimes and Schwarzschild interior}
In this part we the summarize results on understanding of properties of quantum difference equation, including von Neumann stability issues for loop quantization of spacetimes beyond isotropy assumption. As discussed earlier, in the early stages of investigations in LQC, a rigorous formulation of quantum theory, including the physical inner product,  was not available, and resulting physics was not well understood. In absence of these tools,  various early works used notions of pre-classicality at large volumes to understand properties of quantum difference equation in relation to GR. Pre-classicality was obtained in different ways, depending on the procedure of taking  underlying limit, whether of discreteness parameter \cite{bd1}, or to the small spacetime curvature \cite{ck1,ck2,ck3}.
In the above approach, usage of generating function technique was particularly useful to obtain pre-classical solutions analytically \cite{ckb}. With generating functions method, Cartin and Khanna \cite{ck1} showed  non-existence of pre-classical solutions for an early quantization of vacuum Bianchi-I model proposed by Bojowald \cite{mb1}. It was then argued by Date that if one does not require pre-classical solution to correspond to the sector with a vanishing volume, then pre-classical solutions can exist \cite{date}. Cartin and Khanna then studied this model using  separation of variables and investigated existence of
pre-classical solution for all values of triads.  It was found that existence of such pre-classical solutions depends on the sign of the separation parameter \cite{ck2}. It should be noted that presence of matter or cosmological constant may alleviate the problem of pre-classical solutions, as was shown in the analysis of Ref. \cite{khanna_connors}. For the same quantization in the case of   vacuum Bianchi-I LRS model a von Neumann stability analysis with a
symmetric as well a non-symmetric forms of the quantum constraint revealed instabilities in large regions of spacetime \cite{jung-khanna}. In contrast, the quantum difference equation for an earlier quantization of Bianchi-I vacuum spacetime \cite{chiou}, written in the separable form turns out to be stable \cite{gaurav}. Generating functions were also used to study pre-classical solutions in the Schwarzschild interior and limitations of the quantization proposed in Ref. \cite{bh} were identified \cite{ck3}. A von Neumann stability analysis of the same quantization of Schwarzschild interior shows instabilities \cite{jung-khanna}, whereas a different quantization of Schwarzschild interior heuristically motivated by Boehmer and Vandersloot \cite{bh2} demonstrates stability \cite{ckb2}. Nelson and Sakellariadou, studied von-Neumann stability of the quantum difference equation for Bianchi-I vacuum spacetime resulting from the analysis of Ashtekar and Wilson-Ewing \cite{awe2}, and claimed that in the explicit implementation, the difference equation is unstable \cite{nelson3}. As we have discussed in Sec. IVA and above, it can be naive to extend conclusions from one particular choice of matter (or absence of it) to general matter. It is quite possible that in presence of massless scalar the quantum difference equation turns out to be stable \cite{abhay}. Since some of the detailed  properties of the quantum theory, such as self-adjointness of quantum constraint operator and the associated inner product, in above Bianchi-I model remain to be fully understood, it is also possible that instabilities correspond to unphysical solutions which are weeded out in the physical Hilbert space. The effect of implementation of quantum difference equation in an implicit scheme, or by use of different variables remains to be understood.

We now discuss an important technical issue which arises in the quantization of Bianchi spacetimes and the Kantowski-Sachs model. In these models  quantum constraint operator in LQC is a difference equation in more than one variables. Instead of an ordinary difference equation as in the case of isotropic models, one has to solve a partial difference equation. This partial difference equation can not be expressed in a form such that it is  uniformly discrete in all the  variables. Using appropriate redefinition of variables, it has been shown that the partial difference equation can be made uniformly discrete in one of the variables (see for eg. Ref. \cite{ckb2}, and Ref. \cite{awe2} for a rigorous implementation at the quantum theory level in Bianchi-I model). Due to the non-uniform step size, the structure of difference equation is such that the value of the wavefunction at some future discrete step can not be recursively obtained from previous iterations. Further, the discretization step in one direction can also depend on the one in the other direction, and the partial difference equation is not variable separable.

To overcome these problems, Sabharwal and Khanna have recently used a local interpolation method \cite{khanna_sabh}, and Nelson and Sakellariadou have proposed a Taylor expansion scheme for interpolation \cite{nelson2}. We illustrate below the difficulty of non-uniform grids and summarize the interpolation method of Ref. \cite{khanna_sabh} in a simple case where separability in variables is allowed.   Let us assume a quantum constraint with two discrete variables $(\delta, \tau)$, which are separable, and the partial difference equation  can be written as two coupled difference equations. If $C_i$ (where $i = 0,1,2)$ denote coefficients in the $\tau$ difference equation, then say at $\delta = \delta_o$, recursive relation can be used to determine $C_2$, given $C_o$ and $C_1$ as initial data. However, due to non-uniform stepping, at $\delta = \delta_1$, the recursive relation in $\tau$ would have coefficients which depend on value of $\tau$ not obtained from uniform stepping of the values at $\delta = \delta_o$, such as $C_{2 - \epsilon}, C_2$ and $C_{2 + \epsilon}$. Using difference equations, it is not possible to determine  $C_{2 - \epsilon}$ from $C_2, C_1$ and $C_o$, and hence one can not use recursive relation at $\delta = \delta_1$ to obtain $C_{2 + \epsilon}$. In Sabharwal and Khanna's algorithm, one performs a least square fit at $\delta=\delta_o$ to obtain a locally accurate formula which is used to
determine $C_{2 - \epsilon}$ from $C_2, C_1$ and $C_o$. This allows to compute $C_{2 + \epsilon}$ at $\delta = \delta_1$ using difference equation at $\delta = \delta_1$. This value is used to obtain a revised locally accurate formula in the neighborhood of $\tau = 2 - \epsilon, 2, 2 + \epsilon$, which is then used to compute coefficients in the next step $\delta = \delta _2$. The local interpolation method is applicable in the regime where solutions are varying slowly in the neighborhood. Due to similar behavior of solutions at small spacetime curvature (or large volumes) in various models, above method is useful to understand large volume behavior of quantum difference equations, von Neumann stability and evolution of semi-classical states at late times. However, in the regime where solutions can change rapidly in the neighboring points, a more general method, such as the one based on Taylor series expansion for interpolation on neighboring points is expected to be more reliable, as it requires only analyticity of the solution and provides a better control on the accuracy of method \cite{nelson2}. At this stage, more  work is needed to understand the domain of applicability of these methods.

\begin{figure}[tbh!]
\includegraphics[angle=270,width=0.7\textwidth]{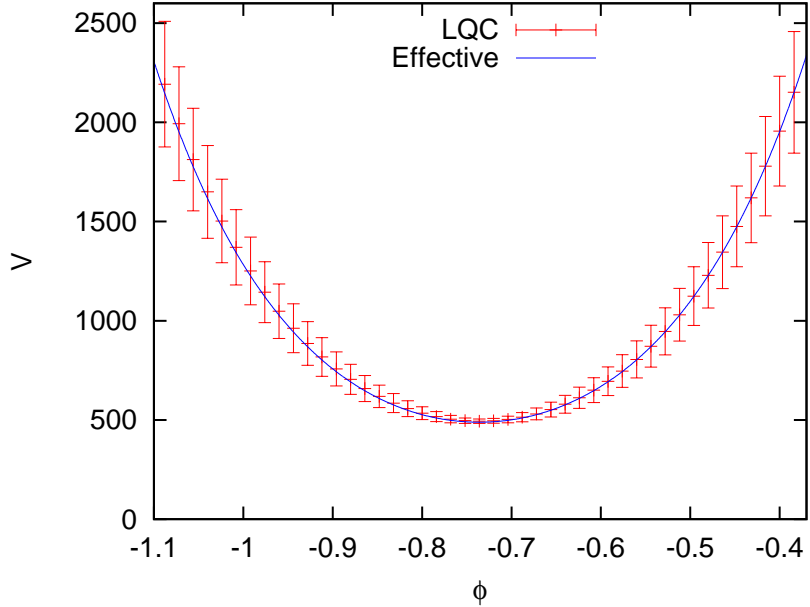}
\caption{This plot demonstrates the validity of effective dynamical trajectory for the massless scalar model in LQC for the numerical simulation of quantum evolution of initial state discussed in Fig. 2. The expectation values of volume observable are peaked on the effective trajectory (solid line) obtain from eq.(\ref{effham}) throughout the evolution.}
\end{figure}

\section{Effective dynamics}
In the previous sections we have discussed different numerical methods to understand detailed features of quantum difference equation in LQC. Interestingly, various aspects of quantum dynamics in LQC can be understood also in terms of
an effective spacetime description using effective dynamical differential equations derived from an effective Hamiltonian constraint. This description is derived in LQC using geometric formulation of quantum mechanics where one treats the space of quantum states as an infinite dimensional quantum phase space with a symplectic structure and relates it to the classical phase space by seeking a faithful embedding \cite{aa_ts}. Such an embedding can be obtained by a judicious choice of states, such as coherent states which at late times peak on classical trajectories in a macroscopic universe. In LQC, an effective Hamiltonian up to quantum corrections corresponding to properties of the states such as fluctuations, has so far been achieved for the  spatially flat model in presence of a massless scalar \cite{vt} and dust \cite{jw}, and generalization for arbitrary matter is in progress \cite{psvt}. Interestingly, using extensive numerical simulations on evolution of states with quantum difference equations in LQC,
validity of effective Hamiltonian has been verified for models in the presence of spatial curvature \cite{apsv,kv}, inflationary potential \cite{aps4} and even anisotropies \cite{szulc_b1,madrid_bianchi1}.\footnote{Another approach which can be used to obtain an effective Hamiltonian is a truncation method \cite{mbeff}. This method is also based on geometric formulation of quantum mechanics, and is similar to the approach of order by order perturbative analysis. It allows a discussion of broader set of state, however extra care is needed to systematically control truncation errors. This method has not been as widely tested in LQC models as the one based on embedding approach. For more discussion of analysis based on this approach we refer the reader to Refs. \cite{mbeff2,mbeff3}.} These works, hence, strongly indicate that the validity of effective Hamiltonian approach in LQC is more general. We illustrate the comparison of effective dynamics with quantum evolution in LQC for the numerical simulation discussed in Fig. 2, in Fig. 4. It can be clearly seen that effective dynamical trajectory obtained from the effective Hamiltonian constraint discussed below (eq.(\ref{effham})) is in an excellent agreement with the expectation values of volume observable in LQC computed by state evolution, at all scales. In a recent work, this agreement has been observed even for states which bounce at very small volume \cite{dgs}.

In LQC, effective dynamics has yielded many insights on the new physics at Planck scale, including issues ranging from  genericity of singularity resolution \cite{ps09,sv,ps11}, and phenomenological issues such as attractor behavior in inflationary dynamics \cite{svv,sloan,rs} and implications for probability of inflation \cite{sloan,ck_infl}, to
detailed analysis of approach to Planck scales in anisotropic models in LQC \cite{gs2,cm2}, and on inclusion of inhomogeneous degrees of freedom \cite{hybrid} (for discussion of various applications see, Sec. V of Ref. \cite{as}). Various results obtained in effective dynamics rely on a use of
numerical methods to solve coupled ODE's, which carefully incorporate control on error terms, arising for example in  a slight non-vanishing of the Hamiltonian constraint in a numerical simulation. For examples of works with detailed discussion on these issues in LQC, we refer the reader to Refs. \cite{sloan,rs,cm2,gs2}.

Amongst various applications of effective dynamics, a very useful one concerns with its use to gain insights on the problems with certain discretizations in LQC, such as old quantization \cite{mb1,abl} or arbitrary discretization as proposed in lattice refinement scheme \cite{lattice}. Below we discuss this aspect of effective dynamics, and show the way it leads to reveal similar problems with certain quantizations as demonstrated in von Neumann stability analysis. Our discussion is based on the analysis in Ref. \cite{cs08}. Since this part of the manuscript has some parallel with the classical cosmology, we work with the lapse $N=1$. 
The effective Hamiltonian constraint for a spatially flat isotropic model with a massless scalar field in LQC is given by \cite{vt}
\be\label{effham}
 - \f{3 \hbar}{4 \gamma} \, \f{\sin^2(\lambda \b)}{\lambda^2} \, \nu + {\cal H}_\phi \approx 0 ~,
\ee
where ${\cal H}_\phi = p_\phi^2/2 |p|^{3/2}$ is the Hamiltonian for the massless scalar. It is straightforward to verify using eq.(\ref{bnudef}), that in the regime $\lambda \b \ll 1$ above equation yields the classical Hamiltonian constraint (\ref{class_cons}) for lapse $N = 1$. Recall that on classical solutions of GR for massless scalar model, $\b = \gamma H \propto a^{-3}$, and  hence, effective Hamiltonian constraint approximates classical GR at large volumes. The same conclusion is reached for effective Hamiltonian constraint with arbitrary matter satisfying weak energy condition \cite{cs08}. This shows that effective Hamiltonian constraint (\ref{effham}) leads to a consistent infrared behavior. To understand the ultraviolet behavior, we consider the expression for energy density of the scalar field, which
using eq.(\ref{effham}), turns out to be,
\be
\rho = \f{1}{2} \, \f{p_\phi^2}{V^2} = \f{3}{8 \pi G} \f{\sin^2(\lambda \b)}{\gamma^2 \lambda^2} ~.
\ee
The maximum value of energy density is given by $\rmax = 3/8 \pi G \gamma^2 \lambda^2 \approx 0.41 \rho_{\rm{Planck}}$, a universal constant. Recall that this is the same value where bounce occurs in numerical evolution of quantum states in LQC \cite{aps3} and the value which emerges as an upper bound on expectation values for all states in the physical Hilbert \cite{acs}. It turns out that the effective dynamics also predicts a bounce at this energy density, and a period of superinflation between $\rmax < \rho < \rmax/2$ \cite{ps06}, whose many interesting implications have been studied in LQC.

So far we have discussed the effective Hamiltonian constraint in the improved quantization in LQC which  yields a quantum difference equation which is uniformly discrete in volume. We now discuss some features of the effective dynamics in cases when quantization leads to a difference equation not uniformly discrete in volume. Examples are the cases of old quantization in LQC where the quantum difference equation is uniformly discrete in triad eigenvalues \cite{aps2}, and lattice refinement scheme where arbitrary discretizations are allowed \cite{lattice}. All these cases can be studied by considering phase space variable $P_m = c |p|^m$, for $-1/2 < m < 0$ (introduced in Sec. IIA), and its conjugate $x = |p|^{1-m}/(1 - m)$. For a quantization based on these variables, the expression for energy density from the effective Hamiltonian constraint, which is similar to (\ref{effham}) with sinusoidal term given by $\sin^2(\lambda_{P_g} P_g)$, turns out to be \cite{cs08}
\be
\rho_{}^{(P_m)} = \f{3}{8 \pi G \gamma^2 \lambda_{P_m}^2} \, \left(\f{8 \pi G}{6} \gamma^2 \lambda_{P_m}^2 p_\phi^2\right)^{(2m + 1)/(2m - 2)}
\ee
where $\lambda_{P_m}$ is a constant related to the area gap in LQC. Note that except for the case of $m = -1/2$ which corresponds to $P_m = \b$ (and a uniform discretization in volume), the maximum value of $\rho$ depends on $p_\phi$. As discussed in Sec. IIIC, this leads to a serious problem because $p_\phi$ is not invariant under the rescaling of the fiducial cell, and the maximum energy density in such quantizations, at which bounce occurs, can be reached at arbitrarily small spacetime curvature by a simple rescaling of the fiducial cell. Unfortunately, even fixing a compact spatial topology such as a 3-torus, with say $V_o = 1$, in these quantizations does not help and problem for a consistent infrared behavior persists. Note that for arbitrary $m$, $P_m$ varies with scale factor on classical trajectories as,
\be
P_m = c p^m = \gamma \dot a \, a^{2m} \propto a^{-(3 w + 1 - 4 m)/2}
\ee
where we have used classical equation $\dot a \propto a^{(-3 w + 1)/2}$ following from classical Friedmann equation and stress-energy conservation law. For  $-1/2 < m \leq 0$, $P_m$ increases with an increase in the scale factor of the universe for  $-1 \leq  w < 0$.
This implies that the term $\sin^2(\lambda_{P_g} P_g)$ in the effective Hamiltonian constraint for $-1/2 < m \leq 0$ leads to large deviations from
GR even for matter which satisfies weak energy condition. This result is in confirmation with our discussion on
problems with old quantization (which corresponds to $m=0$)and arbitrary discretizations ($-1/2 < m < 0$) for von Neumann stability issues of quantum difference equation. In the light of the above argument, in Sec. IVA3 one would have expected that the old quantization of LQC will be problematic in achieving correct classical limit for the case of cosmological constant, which turns out to be true. In fact, it is only for $m = -1/2$ that one obtains a consistent infrared behavior. Hence, out of all possible discretizations, uniform discretization in volume is unique in this sense \cite{cs08}.\footnote{A different analysis based on factor ordering ambiguities results in a similar conclusion \cite{nelson1}.} 

Above discussion shows that effective Hamiltonian approach can be very useful in predicting detailed features of quantum difference operators which can be confirmed by more sophisticated numerical methods, such as von Neumann analysis. In particular, innocuous rescaling of variable under change in fiducial cell can teach us important things. In the same spirit as above, different quantizations in Bianchi-I models have been discussed \cite{cs09}, which show the viability of the recent quantization by Ashtekar and Wilson-Ewing \cite{awe2} in comparison to an earlier quantization by Chiou \cite{chiou}. It will be interesting to see if the analysis of Ref. \cite{cs09} provides useful insights on the  von Neumann stability analysis of Bianchi-I models in LQC with matter.

\section{Summary and Outlook}
In this manuscript we have reviewed various numerical techniques which have been used in LQC to understand the behavior of quantum difference operator and physical states. Since the applied numerical methods are diverse, we used the quantization of massless scalar field in isotropic spacetime as a platform to discuss them in a coherent way.  After providing a brief summary of the classical framework and basics of quantization and how it leads to the quantum Hamiltonian constraint as a difference equation with a uniform discretization in volume, we described the underlying procedure of obtaining physical states using an FFT and with an internal time evolution. These techniques have been extensively used in various isotropic models and also in preliminary investigations of anisotropic models in LQC. In all the models which have been quantized so far in LQC, these methods reveal existence of bounce when the spacetime curvature becomes Planckian. A lot of work in LQC has been devoted to understand the behavior of quantum difference operator at large volumes and consistency of quantization at infrared scales by comparison with the Wheeler-DeWitt theoru using  von Neumann stability analysis. Illustrating these techniques for the isotropic model, we showed the way improved quantization of Ref. \cite{aps3} overcomes problems associated with the old quantization in LQC \cite{mb1,abl,aps2}. Insights on these issues were also discussed using effective dynamics, and we highlighted the complementarity of these approaches to understand infrared issues. It is notable that the unique viable discretization which is established from these works, emerges naturally from loop quantization \cite{aps3}. It will be useful to apply these ideas in greater synergy for anisotropic and black hole spacetimes to gain insights on the properties of quantum difference equations in these models.

Numerical techniques in LQC have so far played a very important role in its progress, however, the field of numerical loop quantum cosmology is still in its infancy. Various avenues remain to be explored in near future which open opportunities for contributions from experts in numerical and computational methods. We have discussed some of the open directions in the manuscript, which include development of techniques to solve partial difference equations with non-uniform discretization carrying forward the works of Refs. \cite{khanna_sabh,nelson2}. These methods will be in particular helpful to uncover detailed physics of anisotropic and black hole spacetimes in LQC. A related arena is to explore the quantum properties of spacetime in the bounce regime, on the lines of numerical techniques developed in classical GR to explore the structure of spacetime on approach to space-like singularities \cite{berger,garfinkle}. Work on these lines is expected to provide deep insights on the way quantum gravity affects mixmaster behavior, and critical phenomena in gravitational collapse. For the Bianchi models, these avenues are being investigated using effective dynamics which already provide an interesting result, that quantum geometric effects lead to Kasner transitions across the bounce in Bianchi-I spacetimes \cite{gs2}. Critical phenomena and Choptuik scaling in gravitational collapse scenarios is also being studied but so far only partial effects from LQG have been incorporated (see for eg. \cite{massgap,husain,kunst}). These works indicate  that underlying quantum geometry may lead to a mass gap in gravitational collapse scenarios \cite{massgap,husain,kunst}.  Whether or not these results are borne out in a rigorous quantization can only be answered by carrying out extensive numerical simulations in near future. Investigations on these lines are also expected to provide us with insights on the phenomena of black hole evaporation. Recently, insights from singularity resolution in LQC were fruitfully used to
gain understanding of this issue for Callan-Giddings-Harvey-Strominger (CGHS) black holes \cite{cghs1}, and extensive numerical analysis revealed new insights on the quantum evaporation of these black holes \cite{cghs2}. Future work on similar lines with Schwarzschild black holes is expected to be a fertile ground for research where numerical methods will play an important role.

Going beyond the isotropic models, or asking more refined questions even in the isotropic models will require adapting algorithms to high performance computation techniques. To illustrate this, let us consider computational costs on an interesting avenue of research -- to test the validity of effective dynamics in extreme regimes (i.e. at volumes comparable to Planck volume) with states which are more general than the ones considered in this manuscript. In the spatially flat isotropic model, note that the characteristic speeds in Wheeler-DeWitt equation (in the $\nu$ representation), to which quantum difference equation in LQC approximates to at large volumes, are proportional to volume. Since the discreteness in volume is fixed by the quantum geometry, this implies that the `time' step to numerically integrate equation in $\phi$ determined by CFL condition is inversely proportional to the volume. Due to this reason, the computational cost of numerically evolving states with wider spreads is much higher with the same techniques as used for sharply peaked states. Note that a wider state, would require the boundary in volume to be considered at a much higher value than the state which is sharply peaked. And, an increase in boundary volume, implies choosing a smaller time step. As an example, for the numerical simulation used in generation of Fig. 2, the outer boundary was taken at $v = 400000$ and the time interval was divided in 300000 steps. The simulation took approximately 45 minutes on a dual core Opteron 280 machine. A naive implementation of the same algorithm for an outer boundary, say at $v = 4 \times 10^{10}$ (which is typically required for certain types of quantum states), increases the computational time by $10^{10}$ hours!  These costs would only grow for asking similar questions in anisotropic models. Certainly, better algorithms harnessing the power of high performance computing are required, which are being developed \cite{dgs}. These algorithms  will provide us with tools to understand robustness of new physics  in the Planck regime and answer fundamental questions on the emergence of continuum spacetime from quantum geometry. Development of these algorithms will also open new avenues to perform extensive numerical evolution of states in anisotropic and black hole models using HPC's in the near future.

\acknowledgments
We are grateful to Tomasz Pawlowski for his seminal contributions in development of numerical techniques in our joint collaborations in LQC. We warmly thank
Daniel Cartin, Peter Diener, Brajesh Gupt and Gaurav Khanna for many insightful discussions on numerical aspects. We also thank Abhay Ashtekar, David Craig and Jorge Pullin for discussions. Numerical simulations discussed in this manuscript were developed and carried out using computational resources of Center for Computation \& Technology at Louisiana State University, and we thank Brajesh Gupt for help with the figures. 
This work is supported by NSF grant PHY1068743.

\end{document}